\documentclass[aps,prc,twocolumn,showpacs,amsmath,amssymb]{revtex4}

\usepackage{graphicx}
\usepackage{dcolumn}
\usepackage{bm}
\usepackage{hyperref}
\usepackage{mathrsfs}
\usepackage{multirow}
\usepackage{rotating}
\usepackage{enumitem}
\usepackage{diagbox}
\usepackage{threeparttable}
\usepackage{booktabs}
\usepackage[hyphenbreaks]{breakurl}

\begin{document}

\title{Quenched $\Lambda$ spin-orbit splitting by relativistic Fock diagram in single-$\Lambda$ hypernuclei}

\author{Shi Yuan Ding}
\affiliation
{Frontiers Science Center for Rare Isotopes, Lanzhou University, Lanzhou 730000, China}
\affiliation
{School of Nuclear Science and Technology, Lanzhou University, Lanzhou 730000, China}

\author{Zhuang Qian}
\affiliation
{Frontiers Science Center for Rare Isotopes, Lanzhou University, Lanzhou 730000, China}
\affiliation
{School of Nuclear Science and Technology, Lanzhou University, Lanzhou 730000, China}

\author{Bao Yuan Sun\footnote{
Corresponding author (email: sunby@lzu.edu.cn)}}
\affiliation
{Frontiers Science Center for Rare Isotopes, Lanzhou University, Lanzhou 730000, China}
\affiliation
{School of Nuclear Science and Technology, Lanzhou University, Lanzhou 730000, China}

\author{Wen Hui Long}
\affiliation
{Frontiers Science Center for Rare Isotopes, Lanzhou University, Lanzhou 730000, China}
\affiliation
{School of Nuclear Science and Technology, Lanzhou University, Lanzhou 730000, China}

\begin{abstract}
We extend the relativistic Hartree-Fock (RHF) theory to study the structure of single-$\Lambda$ hypernuclei. The density dependence is taken in both meson-nucleon and meson-hyperon coupling strengths, and the induced $\Lambda$-nucleon ($\Lambda N$) effective interactions are determined by fitting $\Lambda$ separation energies to the experimental data for several single-$\Lambda$ hypernuclei. The equilibrium of nuclear dynamics described by the RHF model in normal atomic nuclei, namely, the balance between nuclear attractive and repulsive interactions, is then found to be drastically changed in single-$\Lambda$ hypernuclei, revealing a different role of Fock terms via $\Lambda$ hyperon from the nucleon exchange. Since only one hyperon exists in a single-$\Lambda$ hypernucleus, the overwhelmed $\Lambda N$ and $\Lambda\Lambda$ attractions via the Hartree than the $\Lambda\Lambda$ repulsion from the Fock terms require an alternation of meson-hyperon coupling strengths in RHF to rebalance the effective nuclear force with the strangeness degree of freedom, leading to an improved description of $\Lambda$ Dirac mass and correspondingly a systematically reduced $\sigma$-$\Lambda$ coupling strength $g_{\sigma\Lambda}$ in current models as compared to those relativistic mean-field (RMF) approaches without Fock terms. As a result, the effective $\Lambda$ spin-orbit coupling potential in the ground state of hypernuclei is suppressed, and these RHF models predict correspondingly a quenching effect in $\Lambda$ spin-orbit splitting in comparison with the RMF cases. Furthermore, the $\Lambda$ spin-orbit splitting could decrease efficiently by evolving the hyperon-relevant couplings $g_{\sigma\Lambda}$ and $g_{\omega\Lambda}$ simultaneously, where to reconcile with the empirical value the RHF models address a larger parameter space of meson-hyperon couplings.
\end{abstract}


\maketitle

\section{Introductions}

Hypernuclei are of special interest in finite nuclear systems since they allow one to unveil the feature of baryon-baryon interaction with the degrees of freedom beyond nucleons\cite{Yamamoto1994ptp117.361, Gibson1995Phys.Rep.257.349, Epelbaum2009Rev.Mod.Phys.81.1773, Rijken_Nijmegen_2010}. The wealth of information of in-medium baryon-baryon interaction then impact theoretical predictions and comprehension of the deep interior of neutron stars\cite{Madappa_1997, Lattimer_2004, Tolos_neutron-stars_2020, Burgio_neutron-stars_2021}. In recent years, new opportunities arise for hypernuclear physics with the progress of facilities for radioactive ion beams, such as the Japan Proton Accelerator Research Complex (J-PARC)\cite{Sawada_2007}, the Thomas Jefferson National Accelerator Facility (JLab)\cite{Nakamura_2005}, the Facility for Antiproton and Ion Research (FAIR)\cite{Henning_2004} and the High-Intensity Heavy-ion Accelerator Facility (HIAF)\cite{YANG_HIAF_2013, zhou_2018}. As the least strangeness number $S=-1$, relatively more abundant experimental data have been achieved for the single-$\Lambda$ hypernuclei, including $\Lambda$ separation energy and its spin-orbit splitting, in different mass regions\cite{hashimoto_spectroscopy_2006, feliciello_experimental_2015, gal_strangeness_2016}. Via the proposed novel method such as charge-exchange reactions with heavy ion projectiles, it is expected to produce very-neutron-rich hypernuclei and corresponding resonance states with enhanced production rates in the future\cite{FengPRC2020, Saito_2021, Saito2021NovelMF}, which makes it possible to further understand strangeness-bearing baryon-baryon interactions, namely the hyperon-nucleon ($YN$) and the hyperon-hyperon ($YY$), at the various circumstance of nuclear medium.

To study hypernuclei properties in theoretical side, feasible and reliable $YN$ and $YY$ interactions need to be introduced. Within the SU(3) or SU(6) framework and the boson-exchange picture, Nijmengen\cite{Rijken_Nijmegen_1989, Rijken_Nijmegen_1999, cugnon_Nijmegen_2000, vidana_Nijmegen_2001, Rijken_Nijmegen_2006, schulze_Nijmegen_2010, Rijken_Nijmegen_2010, Schulze_Nijmegen_2013, Rijken2019PRC99.044003} and Juelich potentials\cite{Holzenkamp1989NPA500.485, Speth_1994, Speth_1996, haidenbauer_julich_2005} were proposed to produce the realistic two-body $YN$ and $YY$ interactions. Nowadays, the chiral effective field theory in either non-relativistic or covariant framework also made great progress and has been extended to describe the hyperon-nucleon scattering and hyperon masses in the nuclear medium\cite{Savage_ChEFT_1996, Korpa_ChEFT_2001, Beane_ChEFT_2005, polinder_ChEFT_2006, Machleidt_ChEFT_2011, Haidenbauer_ChEFT_2013, Haidenbauer_ChEFT_2016, li_ChEFT_2016, li_ChEFT_2018, Song_ChEFT_2018, Liu_ChEFT_2021}. In addition, the lattice QCD simulations become possible to establish the baryon-baryon interactions as well\cite{Beane_2007, inoue_baryon-baryon_2010, Beane_2012, nplqcd_collaboration_light_2013}. Consequently, there are ab-initio methods or few-body models performed to study hypernuclei structure with the obtained realistic (chiral) interactions, although mainly in light-mass region, such as the (no-core) shell model\cite{Bogner_no-core_2010, Wirth_no-core_2014, Gazda_no-core_2016, Wirth_no-core_2016, Liebig_no-core_2016, Wirth_no-core_2018, Wirth_no-core_2019, Gal_1971, Millener_2001, Millener_2005, Gal_2005, Millener_2008, Millener_2011} and the cluster model\cite{Yamada_Cluster-model_1988, Hiyama_Cluster-model_1999, Cravo_Cluster-model_2002, Shoeb_Cluster-model_2009, Hiyama_Cluster-model_2009}.

The density functional theory has been well developed and been vindicated successfully in describing not only infinite nuclear matter but the single-particle and collective properties of finite nuclei in the almost entire nuclear chart, e.g. see Ref.\cite{Reinhard_1989, Ring_1996, Bender_2003, VRETENAR_HB_2005, Meng_HB_2006, NIKSIC_BMF_2011, Meng_Halo_2015, liang_hidden_2015, doi:10.1142/9872, SHEN1998435, Qian2018} and therein. Thus, it is naturally extended to include the strangeness degree of freedom, triggered by the necessity to investigate the hyperon-involved nuclear force in different mediums and indicate its dependence on several quantities like baryon density and isospin. A various of models of the density functional theory, including the Skyrme-Hartree-Fock\cite{Rayet_Skyrme-Hartree-Fock_1976, Lanskoy_Skyrme-Hartree-Fock_1997, Zhou_Skyrme-Hartree-Fock_2007, Guleria_Skyrme-Hartree-Fock_2012, Hiyama_Skyrme-Hartree-Fock_2014, Zhou_Skyrme-Hartree-Fock_2016, Zhou_Skyrme-Hartree-Fock_2018, Zhou_Skyrme-Hartree-Fock_2020, Zhang_Skyrme-Hartree-Fock_2021}, the relativistic mean-field (RMF)\cite{Brockmann_RMF_1977, Bouyssy_RMF_1981, glendenning_reconciliation_1991, Jennings_RMF_1994, Sugahara_RMF_1994, Vretenar_RMF_1998, Win_RMF_2008, Zhou_RMF_2011, tanimura_description_2012, Zhou_RMF_2014, sun_mean-field_2016, sun_spin_2017, Yao_BRMF_2017, Zhou_BRMF_2017, Xia_BRMF_2017, Tanimura_Cluster_2019, Zhou_RMF_2020, Rong_RMF_2021}, the relativistic Hartree-Fock (RHF)\cite{long_hyperon_2012, hu_effective_2014, li_hypernuclear_2018} and the quark mean-field model\cite{Toki_1998, Shen_QMF_2002, Wang_QMF_2002, Hu_QMF_2014, xing_QMF_2017}, were carried out to study properties of hypernuclei and hyperonic dense matter, where in general their effective $YN$ and $YY$ interactions were adopted by reproducing hyperon separation energies or scattering data, then as an application were used to explore nature of hypernuclear compact stars\cite{weissenborn_hyperons_2012, lopes_hypernuclear_2014, van_dalen_constraining_2014, Hu_mean-field_2014, Fortin_hypernuclear_2020, li_GW190814_2020}.

The relativistic version of density functional theory, namely the covariant density functional (CDF), takes the spin-orbit interaction into account self-consistently via Dirac spinor, which is crucial for the interpretation of the shell structure and to reveal the origins of hidden pseudospin and spin symmetries in atomic nuclei\cite{doi:10.1142/9872, liang_hidden_2015}. However, when applying the CDF approach to hypernuclear systems, further efforts still need to get a rational description of several quantities. It has been revealed that the spin-orbit splitting of $\Lambda$ hyperons is systematically much smaller than that of nucleons\cite{bruckner_spin-orbit_1978, may_observation_1981, may_observation_1983, ajimura_observation_2001, Akikawa_Hypernuclear_2002, Kohri_hypernucleu_2002, MOTOBA200899}. In order to reproduce these experimental and empirical data, a strong $\omega$-tensor coupling could be introduced in $\Lambda$-relevant effective interactions\cite{jennings_dirac_1990, cohen_relativistic_1991, chiapparini_dirac_1991}. Whereas such a tensor coupling is usually missed in nucleon channels when adopting only the Hartree approximation in CDF models, leading to an inconsistent treatment between hyperons and nucleons. Another problem takes place when one predicts the maximum mass of a neutron star by utilizing the obtained equations of state (EoS) for hyperonic dense matter with CDF calculations\cite{long_hyperon_2012, Hu_mean-field_2014, Fortin_hypernuclear_2020}. The appearance of hyperons in the core of stars softens explicitly the EoS at high densities, correspondingly, resulting in the difficulty of the theoretical predictions in complying with the astronomical observations, referred to the hyperon-puzzle\cite{gal_strangeness_2016, Haidenbauer_PUZZLE_2017, VIDANA_puzzle_2013}. Thus, the detailed information on hyperon-involved interaction, especially its in-medium feature with the density and isospin dependence, is essential to clarify these issues.

In recent decades, the CDF approach has been successfully extended by taking into account the exchange diagram of effective two-body interactions. The meson-nucleon coupling strength is performed with a tuned density dependence, which introduces the in-medium effects of nuclear force phenomenologically from the idea of the Dirac Brueckner-Hartree-Fock calculation\cite{Brockmann1992PRL68.3408}. With the inclusion of the Fock terms, the contribution from $\pi$ meson exchange and the nonlocal self-energies\cite{long_evolution_2008, long_non-local_2009, geng_deformed_2020} as well as the tensor part of nuclear force\cite{jiang_nuclear_2015, jiang_self-consistent_2015, zong_relativistic_2018, wang_quantitative_2018, SHEN-tensor-2018} are involved naturally in these well-developed RHF models. Consequently, essential improvements were achieved in characterizing the nuclear structure and nuclear matter properties, such as shell evolutions \cite{long_evolution_2008, wang_tensor_2013, li_magicity_2016}, excitation and decay modes \cite{liang_spin-isospin_2008, liang_isospin_2009, Liang_excitation_2012, niu_-decay_2013, Niu_excitation_2017, Wang-rho-tensor-2020}, novel feature in exotic and superheavy nuclei\cite{Long_Hartree-Fock-Bogoliubov_2010, lu_description_2013, li_superheavy_2014, Li-Pseudospin-orbit-2016}, the nuclear symmetry energy\cite{sun_neutron_2008, long_hyperon_2012, zhao_kinetic_2015,liu_nuclear_2018} and the nucleon effective mass\cite{long_density-dependent_2006, Li-effective-masses-2016}. The progress in the nuclear many-body problem, with either effective or \emph{ab initio} method, then provide solid pillars to illustrate the nature of nuclear structure\cite{SHEN2019103713}, a matter of course to the hypernuclei.

The updated experimental data of nuclear single-particle properties now help us to further refine nuclear structure models, but the faults in reproducing the single-particle energies could lead to incorrect magic numbers and generate the spurious shell closures\cite{wei_pseudo-spin_2020}. Recently, the delicate balance between nuclear attractive and repulsive interactions in the dynamic nuclear medium is realized vital to correct such a problem, demonstrated by the calculations with alternated density dependence of coupling strengths between RMF and RHF Lagrangians\cite{geng_pseudospin_2019}. As a result, the nucleon spin-orbit splitting and the pseudo-spin symmetry restoration are drastically influenced by the Fork terms and the $\rho$-tensor coupling\cite{long_evolution_2008, long_non-local_2009, Liu_New-magicity_2020}. In addition, it is also found that the nuclear thermodynamical properties could be correlated with the in-medium balance of the effective nuclear force, impacting the phase diagram structure of liquid-gas phase transition\cite{yang_nuclear-matter_2019, yang_nuclear-matter_2021}. Therefore, it is expected that the nuclear dynamical equilibrium varies as well in hypernuclear systems due to the involvement of the exchange diagram, which then could play a role in predicting hyperon's separation energies and its spin-particle properties, consequently determining the hyperon-relevant coupling strengths of the RHF models from the experimental data. In fact, the influence of the hyperon-involved physics via the Fock diagram has been unveiled by previous studies on the symmetry energies and the neutron star properties\cite{long_hyperon_2012, sun-Symmetry-energy-2016, li_hypernuclear_2018}. It was found that the nuclear symmetry energy at high densities is suppressed enormously due to the extra hyperon-induced suppression effect originating from the Fock channel, leading to a relatively small predicted value of the neutron star radius\cite{long_hyperon_2012}. While taking further the contribution from $\Delta$-isobars into account, the predicted maximum mass and tidal deformability are compatible with the data extracted from the GW170817 event\cite{Li_delta_2019}.

In view of the capability of the RHF approach in describing the properties of finite nuclei and nuclear matter, therefore, we naturally perform a theoretical extension to study the structure of hypernuclei in this work. As a first step, we now only focus on the case of $\Lambda$ hypernuclei, while their strangeness-bearing effective interactions will be determined by reproducing the experimental data of $\Lambda$ binding energies for several single-$\Lambda$ hypernuclei. Then it is interesting to further investigate the effect of the Fock diagram on the equilibrium of nuclear dynamics and the spin-orbit splitting of hyperons, which is the motivation of this work. In the following, we will introduce the theoretical framework of the RHF approach for $\Lambda$ hypernuclei in Sec.\ref{Theoretical Framework}. In Sec.\ref{Results and Discussion} the results of single-$\Lambda$ hypernuclei within RHF and RMF calculations will be presented and discussed. Finally, a summary is given in Sec.\ref{Summary and Outlook}.

\section{Theoretical Framework}\label{Theoretical Framework}
\subsection{RHF Lagrangian and Hamiltonian with inclusion of $\Lambda$ hyperon}
In this section, the general formalism of the RHF theory will be briefly introduced, and an extension of the energy functional will be performed to include extra the $\Lambda$-hyperon degree of freedom. From the meson-exchange diagram of nuclear force, the bricks of the Lagrangian density for $\Lambda$ hypernuclei consist of the baryon fields ($\psi_B$)---nucleon ($\psi_N$) and hyperon ($\psi_\Lambda$), the isoscalar meson fields---$\sigma$ meson ($\sigma$) and $\omega$ meson ($\omega^\mu$), the isovector meson fields---$\rho$ meson ($\vec{\rho}^{\mu}$) and $\pi$ meson ($\vec{\pi}$), and the photon field ($A^\mu$). Thus, the Lagrangian density for a $\Lambda$ hypernucleus can be expressed as
\begin{align}\label{eq:Lagrangian}
 \mathscr{L} = \mathscr{L}_B + \mathscr{L}_{IS} + \mathscr{L}_{IV} + \mathscr{L}_A + \mathscr{L}_I,
\end{align}
where the terms of free fields read as
\begin{align}\label{eq:LagrangianBMA}
\mathscr{L}_B=&\sum_B\bar{\psi}_B\left(i\gamma^\mu\partial_\mu-M_B\right)\psi_B,\\
\mathscr{L}_{IS}=&+\frac{1}{2}\partial^\mu\sigma\partial_\mu\sigma-\frac{1}{2}m_\sigma^2\sigma^2\nonumber\\
&-\frac{1}{4}\Omega^{\mu\nu}\Omega_{\mu\nu}+\frac{1}{2}m_\omega^2\omega^\mu\omega_\mu,\\
\mathscr{L}_{IV}=&-\frac{1}{4}\vec{R}^{\mu\nu}\cdot\vec{R}_{\mu\nu}+\frac{1}{2}m_\rho^2\vec{\rho}^{\mu}\cdot\vec{\rho}_\mu\nonumber\\
&+\frac{1}{2}\partial^\mu\vec{\pi}\cdot\partial_\mu\vec{\pi}-\frac{1}{2}m_\pi^2\vec{\pi}\cdot\vec{\pi},\\
\mathscr{L}_{A}=&-\frac{1}{4}F^{\mu\nu}F_{\mu\nu},
\end{align}
where the index $B$ ($B'$ later on) represents different baryons (either nucleon $N$ or hyperon $\Lambda$), with its sum $\sum_B$ over neutron $n$, proton $p$ and hyperon $\Lambda$. $M_B$ and $m_\phi$ give the masses of the baryon and mesons ($\phi=\sigma,\omega^\mu,\vec{\rho}^{\mu},\vec{\pi}$), while $\Omega^{\mu\nu}$, $\vec{R}^{\mu\nu}$ and $F^{\mu\nu}$ are the field strength tensors of vector mesons $\omega_\mu, \vec{\rho}^{\mu}$ and photon $A^\mu$, respectively. The interaction between nucleon (hyperon) and mesons (photon) is involved by the Lagrangian $\mathscr{L}_I$,
\begin{align}\label{eq:LagrangianI}
\mathscr{L}_I=\sum_B\bar{\psi}_B&\left(-g_{\sigma B}\sigma-g_{\omega B}\gamma^\mu\omega_\mu\right)\psi_B\nonumber\\
+\bar{\psi}_N&\left(-g_{\rho N}\gamma^\mu\vec{\tau}\cdot\vec{\rho}_\mu-\frac{f_{\pi N}}{m_\pi}\gamma_5\gamma^\mu\partial_\mu\vec{\pi}\cdot\vec{\tau}\right.\nonumber\\
&\left.~+\frac{f_{\rho N}}{2M_N}\sigma^{\mu\nu}\partial_\nu\vec{\rho}_\mu\cdot\vec{\tau}-e\gamma^\mu\frac{1-\tau_3}{2}A_\mu\right)\psi_N.
\end{align}
Here the $\Lambda$ hyperon (namely $\psi_B$ taken as $\psi_\Lambda$), which is charge neutral with isospin zero, participates only in the interactions propagated by the isoscalar mesons. While for the nucleon (here $\psi_B$ taken as $\psi_N$), namely the neutron or proton distinguished by their opposite projection value $\tau_3=1$ or $-1$ of the isospin operator $\vec{\tau}$, the isovector mesons are also in charge. The coupling constants $g_{\phi B}$ ($g_{\phi N}$) and $f_{\phi N}$ determine the strengths of various meson-baryon (meson-nucleon) couplings by means of baryon-density dependent functions to introduce the nuclear in-medium effects phenomenologically \cite{long_density-dependent_2006}.

With the inclusion of Fock diagrams, several strangeness-bearing mesons, such as $K$, $K^*$ and $\kappa$, could participate in the $\Lambda N$ interactions\cite{Holzenkamp1989NPA500.485, Rijken2019PRC99.044003}. As in this work we focus on the ground state properties of single-$\Lambda$ hypernuclei, their contribution could be suppressed by the Fock diagram itself. For instance, it is shown that the effects of $K$ and $K^*$ mesons could be relatively small in single-$\Lambda$ hypernuclei due to their largely canceled contrbituion\cite{BROCKMANN1981365, Brockmann1981PLB104.256}. Therefore, the relevant effect of strangeness degree of freedom is ignored in our current theoretical calculation.

Based on the standard variational principle, one can deduce the corresponding nucleon (hyperon) Dirac equation, meson Klein-Gordon equations and photon Proca equation from the Lagrangian density $\mathscr{L}$,
\begin{align}
  &(i\gamma^{\mu}\partial_{\mu}+M_{B}+\Sigma_{B})\psi_B=0,\label{eq:field equation for baryon}\\
  &(\Box+m_{\sigma}^{2})\sigma=-g_{\sigma N}\bar{\psi}_{N}\psi_{N}-g_{\sigma \Lambda}\bar{\psi}_{\Lambda}\psi_{\Lambda},\label{eq:field equation for sigma}\\
  &(\Box+m_{\omega}^{2})\omega^{\mu}=+g_{\omega N}\bar{\psi}_{N}\gamma^{\mu}\psi_{N}+g_{\omega \Lambda}\bar{\psi}_{\Lambda}\gamma^{\mu}\psi_{\Lambda},\label{eq:field equation for omega}\\
  &(\Box+m_{\rho}^{2})\vec{\rho}^{\mu}=+g_{\rho N}\bar{\psi}_{N}\gamma^{\mu}\vec{\tau}\psi_{N}+\partial_{\nu}\frac{f_{\rho N}}{2M_N}\bar{\psi}_{N}\sigma^{\mu\nu}\vec{\tau}\psi_{N},\label{eq:field equation for rho}\\
  &(\Box+m_{\pi}^{2})\vec{\pi}=+\partial_{\nu}\frac{f_{\pi N}}{m_{\pi}}\bar{\psi}_{N}\gamma^{5}\gamma^{\nu}\vec{\tau}\psi_{N},\label{eq:field equation for pi}\\
  &\partial_{\nu}F^{\nu\mu}=+e\bar{\psi}_{N}\frac{1-\tau_3}{2}\gamma^{\mu}\psi_{N},\label{eq:field equation for photon}
\end{align}\label{eq:field-equation}
where the square box $\Box\equiv\partial_{\mu}\partial^{\mu}$. The baryon self-energy is denoted by $\Sigma_{B}$ in the Dirac equation \eqref{eq:field equation for baryon}, which takes into account the hypernuclear in-medium effects in describing single-particle properties of nucleon or hyperon. With the help of the propagators $D_\phi$ and $D_A$, the meson- and photon-field operators $\varphi(x)$ ($\varphi=\sigma,\omega^{\mu},\vec{\rho}^{\mu},\vec{\pi},A^{\mu}$) in Eqs. (\ref{eq:field equation for sigma}-\ref{eq:field equation for photon}) can be expressed formally as
\begin{subequations}
\begin{align}
\sigma(x) =& -\sum_{B'} \int dx' \bar{\psi}_{B'}(x') \psi_{B'}(x') \mathscr{G}_{\sigma B'}(x') D_{\sigma}(x,x'), \\
\omega^{\mu}(x) =& +\sum_{B'} \int dx' \bar{\psi}_{B'}(x') \psi_{B'}(x') \mathscr{G}_{\omega B'}^\mu(x') D_{\omega}(x,x'), \\
\vec{\rho}^{\mu}(x) =& +\int dx' \bar{\psi}_{N}(x') \psi_{N}(x') \mathscr{G}_{\rho N}^\mu(x') D_{\rho}(x,x'), \\
\vec{\pi}(x) =& -\int dx'\bar{\psi}_{N}(x') \psi_{N}(x') \mathscr{G}_{\pi N}(x') D_{\pi}(x,x'), \\
A^{\mu}(x) =& +\int dx' \bar{\psi}_{N}(x') \psi_{N}(x') \mathscr{G}_{A N}^\mu(x') D_{A}(x,x').
\end{align}\label{eq:field varphi}
\end{subequations}
Here $x$ is four-vector $(t,\bm{x})$. Correspondingly, we define interaction vertices $\mathscr{G}_{\varphi B}(x)$ for a various of meson(photon)-nucleon(hyperon) coupling channels, which for isoscalar $\sigma$ and $\omega$ mesons are represented as
\begin{subequations}\label{eq:vertice for sigmaomega}
  \begin{align}
    \mathscr{G}_{\sigma B}(x) = &+g_{\sigma B}(x),\\
    \mathscr{G}_{\omega B}^\mu(x) = &+g_{\omega B}(x)\gamma^{\mu}.
  \end{align}
\end{subequations}
Apparently, not only nucleons but the $\Lambda$ hyperon can contribute to the isoscalar meson fields. For the rest, namely the isovector mesons and photon fields, it is natural that their interaction vertices connect only to nucleons since the isoscalar and charge-zero nature of $\Lambda$ hyperon,
\begin{subequations}\label{eq:vertice for rhopiA}
  \begin{align}
    \mathscr{G}_{\rho N}^\mu(x) = &+g_{\rho N}(x) \gamma^{\mu} \vec{\tau} + \frac{f_{\rho N}(x)}{2M_{N}} \sigma^{\nu\mu} \vec{\tau} \partial_{\nu}(x),\\
    \mathscr{G}_{\pi N}(x) = &+\frac{f_{\pi N}(x)}{m_{\pi}} \gamma_{5} \gamma^{\nu} \vec{\tau} \partial_{\nu}(x),\\
    \mathscr{G}_{A N}^\mu(x) = &+e\gamma^{\mu}\frac{1-\tau_{3}}{2}.
  \end{align}
\end{subequations}

Starting from the Lagrangian density $\mathscr{L}$ of Eq.\eqref{eq:Lagrangian} again, one can obtain the effective Hamiltonian operator of $\Lambda$ hypernuclei by doing the general Legendre transformation,
\begin{align}
\hat{H}\equiv&~\hat{T}+\sum_{\varphi}\hat{V}_\varphi\notag\\
=&\int dx \sum_{B} \bar{\psi}_{B}(x)(-i\bm{\gamma}\cdot\bm{\nabla}+M_{B}) \psi_{B}(x)\notag\\
&+ \frac{1}{2} \int dx \sum_{B} \sum_{\varphi} \bar{\psi}_{B}(x) \psi_{B}(x) \mathscr{G}_{\varphi B}(x) \varphi(x),
\end{align}
with the operators $\hat{T}$ for the kinetic and $\hat{V}_\varphi$ for the potential energy. Substituting $\varphi(x)$ in Eqs.\eqref{eq:field varphi} into above expression, the potential energy one $\hat{V}_\varphi$ is then described by the two-body interactions mediated by the exchange of mesons, associating with various meson(photon)-nucleon(hyperon) couplings, namely, $\sigma$-S, $\omega$-V, $\rho$-V, $\rho$-T, $\rho$-VT, $\pi$-PV and A-V, see Ref.\cite{geng_deformed_2020} for details. For the isoverctor $\vec{\rho}^{\mu}$ and $\vec{\pi}$ mesons, $\mathscr{G}_{\varphi B}(x) \varphi(x)$ involves the scalar product $\vec{\tau}\cdot\vec{\tau}$ of isospin, while it implies the sum over four-vector index $\mu$ additionally for the vector ones $\omega^\mu, \vec{\rho}^{\mu}$ and $A^\mu$. For the ground-state properties of $\Lambda$ hypernuclei discussed here, the maximum energy difference between occupied states could be small compared to the masses of the exchanged mesons. As a result, the retardation effects, namely the time component of the four-momentum carried by the mesons and photon, are ignored in current RHF approaches as a simplifying assumption\cite{Horowitz1983NPA.399.529, Bouyssy1987PRC36.380}. Correspondingly, the meson (photon) propagators $D_\phi$ ($D_A$) read as
\begin{align}
D_{\phi}(\bm{x},\bm{x}')=\frac{1}{4\pi}\frac{e^{-m_\phi|\bm{x}-\bm{x}'|}}{|\bm{x}-\bm{x}'|},\quad
D_A(\bm{x},\bm{x}')=\frac{1}{4\pi}\frac{1}{|\bm{x}-\bm{x}'|}.
\end{align}
Taking the $\sigma$ field as an example, the potential operator $\hat{V}_{\sigma}$ then becomes
\begin{align}
\hat{V}_{\sigma}=-\frac{1}{2} \sum_{BB'} \iint d\bm{x} d\bm{x}' \left[\bar{\psi}_{B} \mathscr{G}_{\sigma B} \psi_{B}\right]_{\bm{x}} D_{\sigma}(\bm{x},\bm{x}') \left[\bar{\psi}_{B'} \mathscr{G}_{\sigma B'} \psi_{B'}\right]_{\bm{x}'}.
\end{align}
For the isoscalar mesons, the indices $BB'$ represent not only $NN$ but the $N\Lambda$ and $\Lambda\Lambda$ interactions, while for the rest it is natural that only $NN$ channel is contained.

For completeness, one should consider Dirac spinors with both positive and negative energy solutions when second-quantizing baryon field operators. Although there have been efforts within the RHF framework to consider the Dirac sea effect in infinite baryon and meson system\cite{BIELAJEW1984215, Furnstahl1989PRC.40.321, Wasson1990PRC.42.2040}, the renormalization of Dirac sea for finite nuclei and at finite densities is still hardly clarified and solved. As done by previous RHF works for finite nuclei, the contribution of Dirac sea is then ignored here, and the role of sea could partly be involved by adjusting the effective RHF interactions\cite{Brockmann1978PRC18.1510, Horowitz1982PLB109.341, Bouyssy1987PRC36.380}. Within the no-sea approximation, the baryon field operator $\psi_B$ is therefore expanded on the positive-energy set as
\begin{subequations}
  \begin{align}
    \psi_B(x)
    &=\sum_if_i(\bm{x})e^{-i\epsilon_i t}c_i,\label{eq:fi}\\
    \psi_B^\dag(x)
    &=\sum_if^\dag_i(\bm{x})e^{i\epsilon_i t}c^\dag_i.
  \end{align}
\end{subequations}
where $f_i$ is the Dirac spinor, $c_i$ and $c^\dag_i$ are the annihilation and creation operators for a state $i$. In accordance, the energy functional $E$ is obtained by taking the expectation value of the Hamiltonian with respect to a trial ground state $|\Phi_0\rangle$,
\begin{align}
E & = \left\langle\Phi_{0}|\hat{H}| \Phi_{0}\right\rangle  = \left\langle\Phi_{0}|\hat{T}| \Phi_{0}\right\rangle+\sum_{\varphi}\left\langle\Phi_{0}\left|\hat{V}_{\varphi}\right| \Phi_{0}\right\rangle.
\end{align}
In the Hartree-Fock approximation, $|\Phi_0\rangle$ is chosen to be
\begin{align}
\left|\Phi_{0}\right\rangle = \prod_{i=1}^{A} c_{i}^{\dag}|0\rangle, \quad\text{with}\quad \left\langle\Phi_{0}|\Phi_{0}\right\rangle=1,
\end{align}
where $A$ is the mass number of the hypernucleus, and $\left|0\right\rangle$ is the vacuum state. Then the binding energy of a $\Lambda$ hypernucleus is written by
\begin{align}
  E=&\sum_{B}E_{\rm{kin},B} + \sum_{B}(E_{\sigma,B}^D + E_{\omega,B}^D + E_{\sigma,B}^E + E_{\omega,B}^E)\notag\\
   & + E_{\rho,N}+E_{\pi,N}+E_{\rm{e.m.}} + E_{\rm{c.m.}} + E_{\rm{pair}},
  \label{eq:Etot}
\end{align}
where $E_{\rm{kin},B}$ denotes the kinetic energy functional of baryons. $E_{\sigma,B}^D$ and $E_{\omega,B}^D$ correspond to the Hartree terms of the potential energy functional from $\sigma$ and $\omega$, while $E_{\sigma,B}^E$ and $E_{\omega,B}^E$ represent the Fock terms. In addition, the contributions from $\rho$, $\pi$ and $A$ are denoted by $E_{\rho,N}$, $E_{\pi,N}$ and $E_{\rm{e.m.}}$, respectively. The term of $E_{\rm{c.m.}}$ is owing to the center-of-mass correction to the mean field, and $E_{\rm{pair}}$ considers the contribution from nucleon pairing correlations.

In the density-dependent RHF approach, the meson-baryon coupling strengths are regarded as a function of baryon density $\rho_{b}$. The idea of such a treatment as an effective field theory of nuclear many-body system comes from the Dirac Brueckner-Hartree-Fock (DBHF) calculation based on the one-boson-exchange potential, which takes the in-medium effects of nuclear force into account in terms of the density dependence of nucleon self-energies via relativistic G-matrix\cite{Brockmann1992PRL68.3408}. In general, the coupling strengths can be written by
\begin{align}\label{eq:coupling_constants}
g_{\phi B}\left(\rho_{b}\right)=g_{\phi B}(0) f_{\phi B}(\xi) \quad\text{or}\quad
g_{\phi B}\left(\rho_{b}\right)=g_{\phi B}(0) e^{-a_{\phi B} \xi},
\end{align}
where $\xi=\rho_{b}/\rho_{0}$ with $\rho_{0}$ the saturation density of nuclear matter, and
\begin{align}
  f_{\phi B}(\xi)=a_{\phi B}\frac{1+b_{\phi B}(\xi+d_{\phi B})^2}{1+c_{\phi B}(\xi+d_{\phi B})^2}.
\end{align}
In the above expression, $g_{\phi B}(0)$ corresponds to the free coupling constant at $\rho_b=0$.
By fitting the coupling strengths to reproduce the nucleon self-energies from the DBHF calculations as well as the properties of nuclear matter and the selected finite nuclei, it paves an efficient way of modeling the in-medium effects of nuclear force \cite{Typer1999NPA.656.331, long_density-dependent_2006, Giai2010JPG.37.064043, Roca-Maza2011PRC.84.054309}. Correspondingly, the meson-baryon vertex functions become density dependent as well, where two vertices are dressed by their separate density circumstance since two baryons are actually located at different space points.

\subsection{RHF energy functional of spherical $\Lambda$ hypernuclei}

In the following, the description of $\Lambda$ hypernuclei is restricted to the spherical symmetry. Correspondingly, the complete set of good quantum numbers contains the principle one $n$, the total angular momentum $j$ and its projection $m$, as well as the parity $\pi=(-1)^l$ ($l$ is the orbital angular momentum). By taking the quantum number $\kappa$ to denote the angular momentum $j$ and the parity $\pi$, i.e., $\kappa=\pm(j+1/2)$ and $\pi=(-1)^\kappa\text{sign}(\kappa)$, the Dirac spinor $f_i(\bm{x})$ of the nucleon or hyperon in Eq.\eqref{eq:fi} has the following form with spherical coordinate ($r,\vartheta,\varphi$):
\begin{align}
  f_{n\kappa m}(\bm{x}) =  \frac{1}{r} \left(\begin{array}{c}iG_a(r)\Omega_{\kappa m}(\vartheta,\varphi)\\ F_a(r)\Omega_{-\kappa m}(\vartheta,\varphi) \end{array}\right),
\end{align}
where the index $a$ consists of the set of quantum numbers $(n\kappa) = (njl)$, and $\Omega_{\kappa m}$ is the spherical spinor\cite{Varshalovich_quantum_1988}. Meanwhile, the propagators in Eq.\ref{eq:field varphi} can be expanded in terms of spherical Bessel and spherical harmonic functions as
\begin{align}
D_{\phi}(\bm{x},\bm{x}^{\prime}) = \sum_{L=0}^{\infty}\sum_{M=-L}^{L}(-1)^{M}R^{\phi}_{LL}\left( r, r^{\prime}\right) Y_{LM}\left(\bm{\Omega}\right)Y_{L-M}\left(\bm{\Omega}^{\prime}\right),
\end{align}
where $\bm{\Omega}=(\vartheta,\varphi)$, and $R_{LL}$ contains the modified Bessel functions $I$ and $K$ as\cite{Varshalovich_quantum_1988, abramowitz1964handbook}
\begin{align}
  R_{L L}^{\phi}\left(r, r^\prime\right)&=\sqrt{\frac{1}{rr^{\prime}}} I_{L+\frac{1}{2}}\left(m_{\phi}r_{<}\right) K_{L+\frac{1}{2}}\left(m_{\phi}r_{>}\right),\\
  R_{L L}^{A}\left(r, r^\prime\right)&=\frac{1}{2L+1}\frac{r_{<}^{L}}{r_{>}^{L+1}}.
\end{align}

Hence, restricted to the spherical symmetry, these explicit Dirac spinor and propagators are implemented to deduce various components of the hypernuclear energy functional. The baryon's kinetic energy part reads as
\begin{align}
E_{\rm{kin},B}=&\int dr \sum_{a} \hat{j}_{a,B}^2
\begin{pmatrix}
G_{a,B} & F_{a,B}
\end{pmatrix}\notag\\
&\times
\begin{pmatrix}
-\frac{d}{dr}F_{a,B}+\frac{\kappa_{a,B}}{r}F_{a,B}
+ M_{B}G_{a,B} \\ +\frac{d}{dr}G_{a,B}+\frac{\kappa_{a,B}}{r}G_{a,B}-M_{B}F_{a,B}
\end{pmatrix},
\end{align}
where $\hat{j}_{a,B}^2=2j_{a,B}+1$. The $\Lambda$ hyperon, interacting only via the exchange of $\sigma$ and $\omega$ mesons, results in an additional contribution to the potential energy via the isoscalar channel, which is then divided into the direct and exchange terms in the RHF theory. The direct one can be written as
\begin{subequations}
\begin{align}
  E_{\sigma,B}^D&=2\pi\int r^2dr\rho_{s,B}(r)\Sigma_{S,B}^{\sigma}(r),\\
  E_{\omega,B}^D&=2\pi\int r^2dr\rho_{b,B}(r)\Sigma_{0,B}^{\omega}(r).
\end{align}
\label{eq:Edso}
\end{subequations}
Here $\rho_{s,B}$ and $\rho_{b,B}$ define the scalar and baryon density, respectively, which can be calculated by the radial wave function of nucleon or hyperon,
\begin{subequations}
\begin{align}
&\rho_{s,N}\equiv\frac{1}{4\pi r^{2}}\sum_{i=n,p}\sum_{a}\hat{j}^{2}_{a,i}\left[ G^{2}_{a,i}(r)-F^{2}_{a,i}(r)\right],\\
&\rho_{b,N}\equiv\frac{1}{4\pi r^{2}}\sum_{i=n,p}\sum_{a}\hat{j}^{2}_{a,i}\left[ G^{2}_{a,i}(r)+F^{2}_{a,i}(r)\right],\\
&\rho_{s,\Lambda}\equiv\frac{1}{4\pi r^{2}}\sum_{a}\hat{j}^{2}_{a,\Lambda}\left[ G^{2}_{a,\Lambda}(r)-F^{2}_{a,\Lambda}(r)\right],\\
&\rho_{b,\Lambda}\equiv\frac{1}{4\pi r^{2}}\sum_{a}\hat{j}^{2}_{a,\Lambda}\left[ G^{2}_{a,\Lambda}(r)+F^{2}_{a,\Lambda}(r)\right].
\end{align}
\end{subequations}
Then the total baryon density goes to
\begin{align}
\rho_{b}=\rho_{b,N}+\rho_{b,\Lambda}.
\end{align}

The self-energies of nucleon or hyperon include scalar one $\Sigma_{S,B}$ and vector one $\Sigma_{0,B}$, in which the coupling of isoscalar mesons contributes as follows,
\begin{subequations}
\begin{align}
 \Sigma_{S,B}^{\sigma}(r)&\equiv\sum_{B^\prime}\Sigma_{S,BB'}^{\sigma}\notag\\
 &=-g_{\sigma B}(r)\sum_{B^\prime}\int r^{\prime2}dr^\prime g_{\sigma B^\prime}(r^\prime)\rho_{s,B^\prime}(r^\prime)R^{\sigma}_{00}(r,r^\prime),\label{eq:Sigma_S,B}\\
 \Sigma_{0,B}^{\omega}(r)&\equiv\sum_{B^\prime}\Sigma_{0,BB'}^{\omega}\notag\\
 &=+g_{\omega B}(r)\sum_{B^\prime}\int r^{\prime2}dr^\prime g_{\omega B^\prime}(r^\prime)\rho_{b,B^\prime}(r^\prime)R^{\omega}_{00}(r,r^\prime).\label{eq:Sigma_0,B}
\end{align}
\end{subequations}
Such kind of decomposition of the self-energies is non-trivial, since now the direct terms of isoscalar potential in Eq.\eqref{eq:Edso} are separated so that the mechanism of the equilibrium of nuclear dynamics via the Fock diagram can be revealed readily in the next Section.
\begin{align}
  E_{\sigma,N}^D&=E_{\sigma,NN}^D+E_{\sigma,N\Lambda}^D,&
  E_{\sigma,\Lambda}^D&=E_{\sigma,\Lambda N}^D+E_{\sigma,\Lambda\Lambda}^D,\label{eq:E_s_L^D}\\
  E_{\omega,N}^D&=E_{\omega,NN}^D+E_{\omega,N\Lambda}^D,&
  E_{\omega,\Lambda}^D&=E_{\omega,\Lambda N}^D+E_{\omega,\Lambda\Lambda}^D.\label{eq:E_o_L^D}
\end{align}

The contribution of the Fock diagram to the energy functional can be written in a general form as\cite{geng_deformed_2020}
\begin{align}\label{eq:E_LL^E}
E_{\phi,B}^E&=\frac{1}{2}\int drdr^\prime\sum_{a}\frac{\hat{j}_{a,B}^2}{4\pi}
\notag\\
&\times
\begin{pmatrix}
G_{a,B} & F_{a,B}
\end{pmatrix}_{r}
\begin{pmatrix}
Y_{G_{a,B}}^{\phi} & Y_{F_{a,B}}^{\phi}\\
X_{G_{a,B}}^{\phi} & X_{F_{a,B}}^{\phi}
\end{pmatrix}_{r,r^\prime}
\begin{pmatrix}
G_{a,B} \\ F_{a,B}
\end{pmatrix}_{r^\prime}.
\end{align}

For vector mesons, notice that $E_{\phi,B}^E$ should be the sum over their time and space components. For the $\Lambda$-involved part of the exchange (Fock) term, only the $\Lambda$ hyperon itself could take part because only isoscalar couplings remain in $\phi$, namely $E_{\phi,\Lambda}^E=E^{E}_{\phi,\Lambda\Lambda}$. To express the nonlocal self-energies $Y_{G}$, $Y_{F}$, $X_{G}$ and $X_{F}$ in compact form, we introduce several nonlocal densities $\mathcal{R}$ as the source terms,
\begin{subequations}
\begin{align}
\mathcal{R}_{b,B}^{++}(r,r^\prime)=&\hat{j}_{b,B}^2G_{b,B}(r)G_{b,B}(r^\prime),\\
\mathcal{R}_{b,B}^{+-}(r,r^\prime)=&\hat{j}_{b,B}^2G_{b,B}(r)F_{b,B}(r^\prime),\\
\mathcal{R}_{b,B}^{-+}(r,r^\prime)=&\hat{j}_{b,B}^2F_{b,B}(r)G_{b,B}(r^\prime),\\
\mathcal{R}_{b,B}^{--}(r,r^\prime)=&\hat{j}_{b,B}^2F_{b,B}(r)F_{b,B}(r^\prime).
\end{align}
\end{subequations}
Thus for the $\sigma-S$ coupling, the nonlocal self-energies are assembled as follows
\begin{subequations}
\begin{align}
Y_{G_{a,B}}^{\sigma}=&+\sum_{b}\mathcal{J}_{ab}^{\sigma}\mathcal{R}_{b,B}^{++}(r,r^\prime)g_{\sigma B}(r)g_{\sigma B}(r^\prime)\mathcal{D}_{Y_{G}}^{\sigma},\\
Y_{F_{a,B}}^{\sigma}=&-\sum_{b}\mathcal{J}_{ab}^{\sigma}\mathcal{R}_{b,B}^{+-}(r,r^\prime)g_{\sigma B}(r)g_{\sigma B}(r^\prime)\mathcal{D}_{Y_{F}}^{\sigma},\\
X_{G_{a,B}}^{\sigma}=&-\sum_{b}\mathcal{J}_{ab}^{\sigma}\mathcal{R}_{b,B}^{-+}(r,r^\prime)g_{\sigma B}(r)g_{\sigma B}(r^\prime)\mathcal{D}_{X_{G}}^{\sigma},\\
X_{F_{a,B}}^{\sigma}=&+\sum_{b}\mathcal{J}_{ab}^{\sigma}\mathcal{R}_{b,B}^{--}(r,r^\prime)g_{\sigma B}(r)g_{\sigma B}(r^\prime)\mathcal{D}_{X_{F}}^{\sigma},
\end{align}
\end{subequations}
where the isospin factor $\mathcal{J}_{ab}$ is $\delta_{\tau_a\tau_b}$ for the isoscalar channel\cite{Long_Hartree-Fock-Bogoliubov_2010}, and the Clebsch-Gordan coefficients and propagators in nonlocal terms are dealed with
\begin{equation}
\mathcal{D}_{Y_{G}}^{\sigma}=\mathcal{D}_{Y_{F}}^{\sigma}=\mathcal{D}_{X_{G}}^{\sigma}=\mathcal{D}_{X_{F}}^{\sigma}
=\sum_{L}^{\prime}\left(C_{j_a\frac{1}{2}j_b-\frac{1}{2}}^{L0}\right)^2R_{LL}^{\sigma}(r,r^\prime).
\label{eq:C-G_sigma}
\end{equation}
Here it should be noticed that $E^{E}_{\phi,\Lambda\Lambda}$ could act even if there is a unique $\Lambda$ hyperon in a $\Lambda$ hypernucleus, which is not attributed directly to the interplay between different $\Lambda$ hyperons but rather the effect via hyperonic field. Moreover, for the ground state of single-$\Lambda$ hypernuclei where $\Lambda$ occupies the orbit 1$s_{1/2}$, one can further confirms a relation between $E_{\sigma,\Lambda\Lambda}^D$ and $E_{\sigma,\Lambda\Lambda}^E$, namely, $E_{\sigma,\Lambda\Lambda}^E=-E_{\sigma,\Lambda\Lambda}^D/2$, by just considering the fact that the C-G coefficient of Fock terms in Eq.\eqref{eq:C-G_sigma} are $(C_{\frac{1}{2}\frac{1}{2}\frac{1}{2}-\frac{1}{2}}^{00})^2=\frac{1}{2}$.

For the time component of vector ($\omega-V$ and $\rho-V$) couplings, the expressions can be obtained by replacing simply the coupling strength $g_{\phi B}$ and the expansion term $R_{LL}^{\phi}$ of the propagator in above $\sigma-S$ case, as well as reversing the sign of $Y_{G}$ and $X_{F}$.
While for the spatial component of vector couplings, the results need to be regrouped, such as for $\omega-V$,
\begin{subequations}
\begin{align}
Y_{G_{a,B}}^{\omega}=&+\sum_{b}\mathcal{J}_{ab}^{\omega}\mathcal{R}_{b,B}^{--}(r,r^\prime)g_{\omega B}(r)g_{\omega B}(r^\prime)\mathcal{D}_{Y_{G}}^{\omega},\\
Y_{F_{a,B}}^{\omega}=&+\sum_{b}\mathcal{J}_{ab}^{\omega}\mathcal{R}_{b,B}^{-+}(r,r^\prime)g_{\omega B}(r)g_{\omega B}(r^\prime)\mathcal{D}_{Y_{F}}^{\omega},\\
X_{G_{a,B}}^{\omega}=&+\sum_{b}\mathcal{J}_{ab}^{\omega}\mathcal{R}_{b,B}^{+-}(r,r^\prime)g_{\omega B}(r)g_{\omega B}(r^\prime)\mathcal{D}_{X_{G}}^{\omega},\\
X_{F_{a,B}}^{\omega}=&+\sum_{b}\mathcal{J}_{ab}^{\omega}\mathcal{R}_{b,B}^{++}(r,r^\prime)g_{\omega B}(r)g_{\omega B}(r^\prime)\mathcal{D}_{X_{F}}^{\omega},
\end{align}
\end{subequations}
while the corresponding coefficient's terms become
\begin{subequations}
\begin{align}
\mathcal{D}_{Y_{G}}^{\omega}=\mathcal{D}_{X_{F}}^{\omega}
=&\sum_{L}^{\prime\prime}\left[2\left(C_{l_a0l_b 0}^{L0}\right)^2-\left(C_{j_a\frac{1}{2}j_b -\frac{1}{2}}^{L0}\right)^2\right]R_{LL}^{\omega}(r,r^\prime),\\
\mathcal{D}_{Y_{F}}^{\omega}=\mathcal{D}_{X_{G}}^{\omega}
=&\sum_{L}^{\prime\prime}\left(C_{j_a\frac{1}{2}j_b-\frac{1}{2}}^{L0}\right)^2R_{LL}^{\omega}(r,r^\prime),
\end{align}
\end{subequations}
where the prime on the summation $\sum\limits^{\prime}_{L} (\sum\limits^{\prime\prime}_{L})$ indicates that $L+l_a+l_b$ must be even (odd) in order to keep the value nonzero. The value of $L$ is truncated by the coupling of angular momentum naturally. Similarly, we can obtain the energy functional for the spatial components of $\rho-V$ and $A-V$ by replacing the expression with their expansion of propagator and coupling constant. For the case of $\pi-PV$ coupling, the time component contribution drops out because the retardation effect is neglected, and the contribution of spatial components is relatively complicated. Since the inclusion of $\Lambda$ hyperon has nothing to do with the CDF results of isoverctor coupling channels, the details of nucleons' contribution are omitted here and could be found in Ref.\cite{geng_deformed_2020}.

Finally, the last two terms in the total energy functional of Eq.\eqref{eq:Etot} are obtained in the following way. The center-of-mass (c.m.) correction is taken microscopically as
\begin{align}
  E_{\rm{c.m.}}=-\frac{1}{2M_T}\left<{\hat P}_{\rm{c.m.}}^2\right>,
\end{align}
where $M_T=\sum\limits_B M_B=A_N M_N+A_\Lambda M_\Lambda$, and $\left<{\hat P}_{\rm{c.m.}}^2\right>$ is the expection value of the square of $P_{\rm{c.m.}}$, while $P_{\rm{c.m.}}$ is the total momentum operator in the c.m. frame\cite{Bender_correction_2000, ZHAO-Center-of-Mass-2009}. The pairing energy $E_{\rm{pair}}$ is treated under the BCS approximation for the open-shell nuclei, where the finite-range Gogny force is chosen as the pairing interaction with the details following the Refs.\cite{long_new_2004, Long_Hartree-Fock-Bogoliubov_2010, geng_deformed_2020}.

The single-particle (nucleon or hyperon) levels need to be determined by solving the Dirac equation.
Within the RHF theory for spherical nuclei, the radial Dirac equations, i.e., the relativistic Hartree-Fock equations, are expressed as the coupled differential-integral equations. It is convenient to define the total nonlocal self-energy $X_{a,B}$ and $Y_{a,B}$ as\cite{long_pseudo-spin_2006,Long_Hartree-Fock-Bogoliubov_2010}
\begin{align}
\begin{pmatrix}
Y_{a,B}(r) \\ X_{a,B}(r)
\end{pmatrix}
=\int dr^\prime\sum_{\phi}
\begin{pmatrix}
Y_{G_{a,B}}^{\phi} & Y_{F_{a,B}}^{\phi}\\
X_{G_{a,B}}^{\phi} & X_{F_{a,B}}^{\phi}
\end{pmatrix}_{r,r^\prime}
\begin{pmatrix}
G_{a,B} \\ F_{a,B}
\end{pmatrix}_{r^\prime}.
\label{eq:TSEnonlocal}
\end{align}
Therefore, the Dirac equation becomes
\begin{align}
\varepsilon_{a,B}
\begin{pmatrix}
G_{a,B}(r) \\ F_{a,B}(r)
\end{pmatrix} =&
\begin{pmatrix}
M_{B}+\Sigma_+^B(r) & -\frac{d}{dr}+\frac{\kappa_{a,B}}{r}+\Sigma_{T}^{B}(r) \\
\frac{d}{dr}+\frac{\kappa_{a,B}}{r}+\Sigma_{T}^{B}(r) & -\left[M_{B}-\Sigma_-^B(r)\right]
\end{pmatrix}\notag\\
&\times
\begin{pmatrix}
G_{a,B}(r) \\ F_{a,B}(r)
\end{pmatrix} +
\begin{pmatrix}
Y_{a,B}(r) \\ X_{a,B}(r)
\end{pmatrix}.
\end{align}
Here the local self-energies $\Sigma_\pm^B=\Sigma_{0,B}\pm\Sigma_{S,B}$ composed by the vector and scalar terms, and $\Sigma_{T}^{B}$ contains the contribution from the direct terms of tensor part\cite{Long_Hartree-Fock-Bogoliubov_2010}. The scalar self-energy $\Sigma_{S,B} = \Sigma_{S,B}^{\sigma}$, and the time component of the vector one has
\begin{align}
  \Sigma_{0,B}(r) = \sum_{\phi}\Sigma_{0,B}^{\phi}(r)+\Sigma_{R}(r),
\end{align}
where $\phi=\omega,\rho$ for nucleons ($B=N$), and $\phi=\omega$ for $\Lambda$ hyperons ($B=\Lambda$).
In addition, $\Sigma_R$ is the rearrangement term due to the density dependence of the coupling constant, which can be divided into a direct $\Sigma_R^{D}$ and an exchange part $\Sigma_R^{E}$,
\begin{align}
\Sigma_R(r)=\Sigma_{R}^{D}(r)+\Sigma_{R}^{E}(r)
=\sum_{\phi}\left[\Sigma_{R,\phi}^{D}(r)+\Sigma_{R,\phi}^{E}(r)\right].
\end{align}
Here $\Sigma_{R,\phi}^{D}$ and $\Sigma_{R,\phi}^{E}$ contain the summation over all baryons for the isoscalar case of $\phi=\sigma,\omega$, but only over nucleons for the isovector one. For example, the direct term from $\sigma-S$ coupling is shown as
\begin{align}
\Sigma_{R,\sigma}^{D}(r)=\sum_{B}\frac{1}{g_{\sigma B}} \dfrac{\partial g_{\sigma B}}{\partial \rho_{b}}\rho_{s,B}\Sigma_{S,B}^{\sigma}(r).
\end{align}
By introducing the nonlocal self-energy $X_{a,B}^\phi$ and $Y_{a,B}^\phi$ of each meson coupling channel like in Eq.\eqref{eq:TSEnonlocal}, the exchange contribution to the rearrangement term reads
\begin{align}
\Sigma_{R,\phi}^{E}(r) = &\sum_{B}\dfrac{1}{g_{\phi B}} \dfrac{\partial g_{\phi B}}{\partial \rho_{b}}\sum_{a}\frac{\hat{j}_{a,B}^2}{4\pi r^2}
\begin{pmatrix}
G_{a,B}Y_{a,B}^\phi + F_{a,B}X_{a,B}^\phi
\end{pmatrix}_{r}
.
\end{align}

\subsection{Spin-orbit coupling potential of $\Lambda$ hyperon}

To study later the spin-orbit splittings, the radial Dirac equation in the RHF theory could be derived further to get a Schr\"odinger-like equation for the upper component $G_{a,B}$\cite{long_pseudo-spin_2006}. For the $\Lambda$ hyperon, one can obtain
\begin{align}
\begin{split}
\varepsilon_{a,\Lambda}G_{a,\Lambda}=&\left\{-\dfrac{1}{M_{+}}\dfrac{d^{2}}{dr^{2}}-\dfrac{1}{M_{+}}
\left[\mathcal{V}_{\rm{CB},\Lambda}+\mathcal{V}_{\rm{SO},\Lambda}\right.\right.\notag\\
&\left.\left.+\mathcal{V}_{1,\Lambda}\dfrac{d}{dr}+\mathcal{V}_{2,\Lambda}\right]
+\Sigma_{+}^{\Lambda}\right\}G_{a,\Lambda},
\end{split}
\end{align}
where $M_{+} = \varepsilon_{a,\Lambda} + 2M_\Lambda - \Sigma_-^\Lambda$.
The induced items in the formula are defined as follows:
\begin{subequations}
\begin{align}
\mathcal{V}_{\rm{CB},\Lambda}&=-\dfrac{\kappa_{a,\Lambda}(\kappa_{a,\Lambda}+1)}{r^{2}},\\
\mathcal{V}_{\rm{SO},\Lambda}&=-\dfrac{\kappa_{a,\Lambda}}{r}\left[ \dfrac{1}{M_{+}'}\dfrac{d}{dr}M_{+}'+\left( X_{G_{a,\Lambda}}+Y_{F_{a,\Lambda}}\right) \right],\\
\mathcal{V}_{1,\Lambda}&=-\dfrac{1}{M_{+}'}\dfrac{d}{dr}M_{+}'+\left(X_{G_{a,\Lambda}}-Y_{F_{a,\Lambda}}\right),\\
\mathcal{V}_{2,\Lambda}&=-\dfrac{X_{G_{a,\Lambda}}}{M_{+}'}\dfrac{d}{dr}M_{+}'+\dfrac{d}{dr}X_{G_{a,\Lambda}}\notag\\
&~~~~-X_{G_{a,\Lambda}}Y_{F_{a,\Lambda}}-M_{+}Y_{G_{a,\Lambda}}+X_{F_{a,\Lambda}}\left(\Sigma_{+}^{\Lambda}-\varepsilon_{a,\Lambda}\right).
\end{align}
\end{subequations}
Here we introduce $M_{+}'\equiv M_{+}-X_{F_{a,\Lambda}}$. The relevant terms of the nonlocal self-energies are given by sum over all coupling channels, e.g., $X_{F_{a,\Lambda}}=\sum_\phi X_{F_{a,\Lambda}}^\phi$. For the single-$\Lambda$ hypernuclei, the nonlocal self-energies of hyperon could be smaller than the local ones considerably. Hence, if we could neglect them, the above Schr\"odinger-like equation can be reduced to the familiar expression taken in RMF theory\cite{meng_pseudospin_1998, liang_hidden_2015}, shown as
\begin{align}
\begin{split}
  &\left\{-\frac{1}{M_{+}(r)} \frac{d^{2}}{d r^{2}}+\frac{1}{M_{+}^{2}(r)} \frac{d M_{+}(r)}{d r} \frac{d}{d r}+\frac{1}{M_{+}(r)}\frac{\kappa_{a,\Lambda}(\kappa_{a,\Lambda}+1)}{r^{2}} \right.\\
  &~~+\left.\frac{1}{M_{+}^{2}(r)} \frac{d M_{+}(r)}{d r} \frac{\kappa_{a,\Lambda}}{r}+\Sigma_+^\Lambda(r)\right\} G_{a,\Lambda}(r)=\varepsilon_{a,\Lambda} G_{a,\Lambda}(r),\label{eq:SchLike}
\end{split}
\end{align}
where the term which is proportional to $\kappa_{a,\Lambda}(\kappa_{a,\Lambda}+1) = l_{a,\Lambda}(l_{a,\Lambda} + 1)$ corresponds to the centrifugal barrier as compared to the Schr\"odinger equation. The spin-orbit coupling potential of $\Lambda$ hyperon is simplified as
\begin{align}
V_{\rm{SO},\Lambda} = \frac{1}{M_{+}^{2}(r)} \frac{d M_{+}(r)}{d r} \frac{\kappa_{a,\Lambda}}{r},
\label{eq:Vso}
\end{align}
which consequently leads to the spin-orbit splitting in the single-particle spectrum of hyperon.

\section{Results and Discussion}\label{Results and Discussion}
Now we can use the RHF theory to investigate the bulk and single-particle properties of the $\Lambda$ hypernuclei. Especially for the simplest hypernuclear system with single-$\Lambda$, the role of the Fock terms via the extra $\Lambda$-hyperon degree of freedom will be illustrated in this section. For the $NN$ interaction, the RHF effective interactions PKO1\cite{long_density-dependent_2006}, PKO2 and PKO3\cite{long_evolution_2008} are utilized in the calculation, in comparison with the RMF effective functional PKDD\cite{long_new_2004}. The Dirac equation is solved in a radial box size of $R=20$ fm with a step of 0.1 fm. For the open-shell nuclei, the pairing correlation is considered within the BCS method. Here it is treated only for $nn$ and $pp$ pairing with the Gogny interaction D1S\cite{Berger_NPA428_23}, and the blocking effect is taken for the last valence nucleon or $\Lambda$ hyperon. Thus, the analysis of the $\Lambda$-hypernuclei could be performed if the $\Lambda N$ effective interaction is further determined.

\subsection{$\Lambda$ binding energies and $\Lambda N$ effective interaction}

The $\Lambda N$ interaction in recent models is related to determine the coupling strengths between $\sigma$- or $\omega$-meson and $\Lambda$ hyperon. For convenience, the ratio between meson-$\Lambda$ and meson-nucleon couplings $g_{\phi\Lambda}/g_{\phi N}$ is introduced. As the utilized RHF $NN$ effective interactions are density dependent, the $\sigma$-$\Lambda$ and $\omega$-$\Lambda$ coupling strengths thus evolve with the baryon density as well. The mass of $\Lambda$ hyperon takes $M_\Lambda = 1115.6 $ MeV. The isoscalar-vector coupling strength $g_{\omega\Lambda}/g_{\omega N}$ is fixed to be 0.666 according to the n\"{a}ive quark model\cite{dover_hyperon-nucleus_1984}. Then the isoscalar-scalar coupling strength $g_{\sigma\Lambda}/g_{\sigma N}$ is adjusted to reproduce the experimental $\Lambda$ binding energies (separation energies) $B_\Lambda$ assuming $\Lambda$ in the 1$s_{1/2}$ state of hypernuclei $^{16}_\Lambda$O, $^{40}_\Lambda$Ca, and $^{208}_\Lambda$Pb\cite{gal_strangeness_2016}, which theoretically $B_\Lambda$ is obtained by the energy difference
\begin{align}
    B_\Lambda(^{A}_\Lambda Z) = E(^{A-1}Z) - E(^{A}_\Lambda Z),
\end{align}
where the binding energy $E$ of the referred nuclei is gained from Eq.\eqref{eq:Etot}.

\begin{table}[hbpt]
  \caption{The $\sigma$-$\Lambda$ coupling strengths $g_{\sigma\Lambda}/g_{\sigma N}$ fitted for CDF effective interactions by minimizing the root-mean-square deviation $\Delta$ (in MeV) from the experiment values of $\Lambda$ binding energies of $^{16}_\Lambda$O, $^{40}_\Lambda$Ca and $^{208}_\Lambda$Pb, where the $\omega$-$\Lambda$ coupling is fixed to be $g_{\omega\Lambda}/g_{\omega N} = 0.666$.}\label{Tab:Coupling Constants}
  \renewcommand{\arraystretch}{1.5}
  \setlength{\tabcolsep}{5pt}
  \begin{tabular}{ccccccccc}
  \hline
       & PKO1-$\Lambda1$ & PKO2-$\Lambda1$ & PKO3-$\Lambda1$ & PKDD-$\Lambda1$\\\hline
  $g_{\sigma\Lambda}/g_{\sigma N}$ & 0.596 & 0.591 & 0.594 & 0.620 \\
  $\Delta$ &0.265 &0.260 &0.407 &0.347 \\
  \hline
  \end{tabular}
\end{table}

For the selected RHF and RMF functionals, the $\sigma$-$\Lambda$ coupling strengths $g_{\sigma\Lambda}/g_{\sigma N}$ is given in Table.\ref{Tab:Coupling Constants}, by minimizing the root-mean-square deviation $\Delta$ for the $\Lambda$ binding energies between theoretical calculation and experimental values,
\begin{equation}
\Delta\equiv\sqrt{\frac{1}{N}\sum\limits_{i=1}^N(B_{\Lambda,i}^{\rm{exp.}}-B_{\Lambda,i}^{\rm{cal.}})^2}.
\end{equation}
The induced $\Lambda N$ effective interactions are named by $\Lambda 1$. All of them reproduce the data of $^{16}_\Lambda$O, $^{40}_\Lambda$Ca, and $^{208}_\Lambda$Pb well with comparable deviation $\Delta$, while PKO2-$\Lambda1$ gives the best agreement in all. With the fixed $g_{\omega\Lambda}/g_{\omega N}$, it is seen from the table that the ratios of $g_{\sigma\Lambda}/g_{\sigma N}$ in RHF are slightly smaller than PKDD's value in RMF. In fact, the values of $g_{\phi N}$ have been found to be dropped systematically in RHF with the inclusion of the Fock terms\cite{long_shell_2007, geng_pseudospin_2019}. Thus, the absolute values of $g_{\phi \Lambda}(\rho_b)$ in three RHF models, namely, $g_{\sigma\Lambda}$ and $g_{\omega\Lambda}$, are sizably suppressed at various baryon density $\rho_b$ than one in RMF, so that the nuclear in-medium balance and the single-particle properties could be affected by the additional hyperon in hypernuclei.

\begin{figure}[hbpt]
 \centering
 \includegraphics[width=0.48\textwidth]{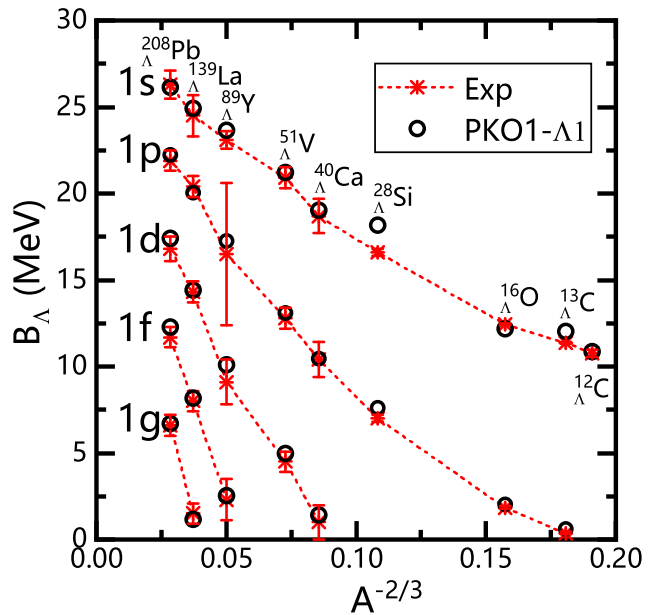}
 \caption{(Color Online) The calculated $\Lambda$ binding energies $B_\Lambda$ for the single-$\Lambda$ hypernuclei with the RHF effective interaction PKO1-$\Lambda1$ in comparison with the experimental data taken from Refs. \cite{hashimoto_spectroscopy_2006, gal_strangeness_2016}.}\label{Fig:B_PKO1}
\end{figure}

Except for the $\Lambda$ binding energies of $^{16}_\Lambda$O, $^{40}_\Lambda$Ca, and $^{208}_\Lambda$Pb, the obtained CDF functionals are also checked to be suitable for describing several other $\Lambda$ hypernuclei with experimental data, where $\Lambda$ can occupy either in ground state 1$s_{1/2}$ or excited states with higher angular momentum $l_\Lambda$. Taking PKO1-$\Lambda1$ as an example, the $B_\Lambda$ result is shown in Figure.\ref{Fig:B_PKO1}, with a relatively large mass range from $^{12}_\Lambda$C to $^{208}_\Lambda$Pb. Since the lack of information on the spin status of $\Lambda$ during the experiments, the calculation performed here just assume that $\Lambda$ occupies the spin-paralleled state, i.e., $j_\Lambda=l_\Lambda+1/2$. It is seen the theoretical predictions in recent models are consistent quite well with the experiments\cite{gal_strangeness_2016}, for both cases of ground and excited hyperon. Besides, we also checked the single-$\Lambda$ potentials $U_\Lambda$ at the saturation baryon density in symmetric nuclear matter, which give $-30.4$ MeV, $-30.2$ MeV and $-30.9$ MeV for PKO1-$\Lambda$1, PKO2-$\Lambda$1 and PKO3-$\Lambda$1, respectively, in agreement with the empirical value such as from \emph{ab initio} calculations\cite{PhysRevC.81.035803}.

To explore the origin of why the $\sigma$-$\Lambda$ coupling ratio $g_{\sigma\Lambda}/g_{\sigma N}$ and its absolute strength $g_{\sigma\Lambda}$ are systematically reduced in RHF calculations, it is deserved to comprehend from a viewpoint of the dynamic nuclear medium. Recently, such kind of method has been carried out successfully to elucidate the mechanism of the pseudo-spin symmetry, the shell evolution and the liquid-gas phase transition\cite{geng_pseudospin_2019, Liu_New-magicity_2020, wei_pseudo-spin_2020, yang_nuclear-matter_2021}. With different treatments of the density dependence of coupling strengths, the delicate balance between nuclear attraction and repulsion in medium could change enormously. For $\Lambda$ hypernuclei, the dynamical equilibrium effects in nucleon's and $\Lambda$'s channel couple each other and interplay via the meson-exchange. It is much easier to enlighten these effects within a light single-$\Lambda$ hypernucleus due to the relatively larger ratio of hyperon to nucleon numbers than those in heavier hypernuclei. With even numbers of proton and neutron in nucleon part, therefore, the single-$\Lambda$ hypernucleus $^{13}_\Lambda$C can be taken as a good example to proceed by separating its energy functional with various components, as organized in Table.\ref{Tab:EnergyFunctional}. Here $E_{\rm{others}}$ includes the energy contribution from the isovector meson-nucleon couplings, the Coulomb field, the pairing and the center-of-mass corrections.

\begin{table*}[hbpt]
  \caption{The kinetic and potential energies (in MeV) in various channels for the single-$\Lambda$ hypernucleus $^{13}_\Lambda$C. The results are given by the RHF effective interactions PKO1-$\Lambda1$, PKO1-$\Lambda1'$ and the RMF one PKDD-$\Lambda1$. Specifically, the terms of $E_{\phi,\Lambda\Lambda}$ correspond to the potential energy originated from $\Lambda\Lambda$ interaction, see Eqs. \eqref{eq:E_s_L^D}, \eqref{eq:E_o_L^D} and \eqref{eq:E_LL^E} for the detail.}
  \label{Tab:EnergyFunctional}
  \renewcommand{\arraystretch}{1.5}
  \doublerulesep 0.1pt \tabcolsep 7pt
  \begin{tabular*}{\textwidth}{crrrcrrrcrrr}
  \hline
   \multirow{2}{*}{$^{13}_\Lambda$C} & \multicolumn{3}{c}{PKO1-$\Lambda1$} && \multicolumn{3}{c}{PKO1-$\Lambda1'$} && \multicolumn{3}{c}{PKDD-$\Lambda1$} \\
   \cline{2-4} \cline{6-8} \cline{10-12}
    & Nucleon & Lambda & ($E_{\phi,\Lambda\Lambda}$) && Nucleon & Lambda & ($E_{\phi,\Lambda\Lambda}$) && Nucleon & Lambda & ($E_{\phi,\Lambda\Lambda}$)  \\
  \hline
     $E_{\rm{kin}}$   &      200.88  &      7.90  &       $-$                 &&      202.49  &      9.80  &       $-$    &&      184.51  &      8.18  &       $-$             \\
     $E_\sigma^D$     &    -1344.62  &    -72.73  &     (-3.83)               &&    -1377.27  &    -86.14  &     (-5.44)  &&    -1694.45  &    -99.84  &     (-5.86)           \\
     $E_\omega^D$     &     1003.35  &     59.47  &      (3.44)               &&     1029.76  &     68.23  &      (4.60)  &&     1415.49  &     87.64  &      (5.43)           \\
     $E_\sigma^E$     &      284.30  &      1.92  &      (1.92)               &&      288.56  &      2.72  &      (2.72)  &&        0.00  &      0.00  &      (0.00)           \\
     $E_\omega^E$     &     -200.23  &     -1.68  &     (-1.68)               &&     -203.18  &     -2.23  &     (-2.23)  &&        0.00  &      0.00  &      (0.00)           \\
     $E_{\rm{others}}$&      -39.27  &     -0.73  &       $-$                 &&      -39.61  &     -0.92  &       $-$    &&       -0.41  &     -0.77  &       $-$             \\
  \hline
   \multirow{2}{*}{Sum} & -95.59 & -5.85 &&& -99.25 & -8.54 &&& -94.86 & -4.79 \\
    \cline{2-3} \cline{6-7} \cline{10-11}
    &\multicolumn{2}{c}{-101.44} &  && \multicolumn{2}{c}{-107.79} &  && \multicolumn{2}{c}{-99.65}  \\
    \hline
  \end{tabular*}
\end{table*}

As one can see from Table.\ref{Tab:EnergyFunctional}, the binding energy is dominated by the cancelation between strong attraction ($E_\sigma$) and strong repulsion ($E_\omega$) from the isoscalar meson coupling channels\cite{geng_pseudospin_2019}. First, let's compare the values of the "Nucleon" channel with RHF functional PKO1-$\Lambda1$ to RMF one PKDD-$\Lambda1$. Because of the limited $\Lambda$ number as well as the isospin difference between hyperon and nucleon, the $\Lambda$ hyperon-induced mean field (or self-energy) in a single-$\Lambda$ hypernucleus impacts little on the nucleon field, and could be regarded as a perturbation effect. Thus, the dynamical equilibrium in the nucleon channel is dominated by the nucleons themselves. As a result, RHF provides a stronger residual attraction given by $E_\sigma^D+E_\omega^D$ from the direct isoscalar terms than RMF, which help to cancel the extra repulsion $E_\sigma^E+E_\omega^E$ introduced by the exchange diagram and correspondingly persist in the balance of nucleons.

However, the situation in the hyperon channel (columns marked by ``Lambda" in Table.\ref{Tab:EnergyFunctional}) is drastically changed. The inclusion of $\Lambda$ hyperon inside a normal atomic nucleus results in an extra attractive potential. Despite only one hyperon, $\Lambda$ can contribute a distinct value of the binding energy via the direct terms of isoscalar meson coupling, i.e., $E_\sigma^D+E_\omega^D$, because there are not only $\Lambda$ itself but many nucleon friends participating indeed to dress its self-energies as shown in Eqs.\eqref{eq:Sigma_S,B} and \eqref{eq:Sigma_0,B}. On the contrary, the contribution from the Fock terms to $\Lambda$-relevant energy functionals is suppressed critically since only $\Lambda$ itself accounts for the nonlocal self-energy and then $E_{\phi,\Lambda}^E=E^{E}_{\phi,\Lambda\Lambda}$, seeing the values of $E_{\phi,\Lambda\Lambda}$ in the table. If one defines a relative ratio of the isoscalar potential energy between Fock and Hartree channels, written as
\begin{equation}
\chi\equiv\left|\frac{E_\sigma^E+E_\omega^E}{E_\sigma^D+E_\omega^D}\right|,
\end{equation}
it is seen that $\chi_N\thickapprox24.6\%$ and $\chi_\Lambda\thickapprox1.8\%$ for PKO1-$\Lambda1$. Therefore, in the case of single-$\Lambda$ hypernuclei, the equilibrium of nuclear dynamics in the hyperon channel is controlled mainly by the direct isoscalar terms, showing a very different mechanism from the nucleon channel. Consequently, the change of the $\sigma$-$\Lambda$ coupling strength $g_{\sigma\Lambda}/g_{\sigma N}$ from PKDD-$\Lambda1$ to PKO1-$\Lambda1$ could be indicated by comparing the details of their energy functionals.

To clarify separately the role of meson-nucleon and meson-hyperon coupling strengths on the hypernuclear binding energy, it is helpful to introduce a tentative CDF functional named by PKO1-$\Lambda1'$, which adopts the RHF effective interaction PKO1 to give the meson-nucleon coupling strengths but takes the value of $g_{\sigma\Lambda}/g_{\sigma N}=0.620$ in PKDD to determine the hyperon's contribution. So the energy differences between PKDD-$\Lambda1$ and PKO1-$\Lambda1'$ given in Table.\ref{Tab:EnergyFunctional} could ascribe to the alternated meson-nucleon coupling strengths from RMF to RHF, while the deviations of PKO1-$\Lambda1'$ from PKO1-$\Lambda1$ are associated with the change of $g_{\sigma\Lambda}/g_{\sigma N}$.

It is shown that in PKO1-$\Lambda1'$ the hypernucleus becomes more binding owing to both nucleon and $\Lambda$ parts in comparison with PKDD-$\Lambda1$, leading to a bad description to reproduce the observed $\Lambda$ separation energy within PKO1-$\Lambda1'$. In fact, the CDF potential energy could be divided into $E_{\phi,NN}$, $E_{\phi,N\Lambda}$, $E_{\phi,\Lambda N}$ and $E_{\phi,\Lambda\Lambda}$ according to the type of interacting particles, and $E_{\phi,N\Lambda}=E_{\phi,\Lambda N}$. It is checked that the total contribution from nucleons themselves, which includes nucleons' kinetic energy and the potential via $NN$ channel, is robust enough against the change of meson-nucleon coupling strengths from RMF to RHF functional, due to the balance from the extra exchange diagram. But the $N\Lambda$-relevant terms differ remarkably, which give $E_{\sigma,N\Lambda}+E_{\omega,N\Lambda}=-11.78$ MeV in PKDD-$\Lambda1$ and $-17.07$ MeV in PKO1-$\Lambda1'$. Therefore, as both affected by $E_{\phi,N\Lambda}$, not only the $\Lambda$ binding is enhanced from $-4.79$ to $-8.54$ MeV, but the binding from nucleon channels reinforces with the $\Lambda$ polarization effect, seeing the Eq.\eqref{eq:E_s_L^D} and \eqref{eq:E_o_L^D}. The failed description of hypernuclear binding energy in RHF models by using the RMF's value of $g_{\sigma\Lambda}/g_{\sigma N}$ implies that the density dependence of meson-hyperon coupling strengths may differ tangibly from meson-nucleon ones, additionally deviate between $g_{\sigma\Lambda}$ and $g_{\omega\Lambda}$.

\begin{table}[hbpt]
  \caption{The similar to Tab. \ref{Tab:EnergyFunctional} but for the contributions to the $\Lambda$ binding energy $B_\Lambda$ (in MeV) of the single-$\Lambda$ hypernucleus $^{13}_\Lambda$C from various channels of corresponding energy density functional.}
  \label{Tab:Binding Energies}
  \renewcommand{\arraystretch}{1.5}
  \setlength{\tabcolsep}{4pt}
  \begin{tabular}{crrcrrcrr}
  \hline\hline
  \multirow{2}{*}{$B_\Lambda$} & \multicolumn{2}{c}{PKO1-$\Lambda1$} && \multicolumn{2}{c}{PKO1-$\Lambda1'$} && \multicolumn{2}{c}{PKDD-$\Lambda1$} \\
  \cline{2-3} \cline{5-6} \cline{8-9}
                               & N   & $\Lambda$    && N  & $\Lambda$    && N  & $\Lambda$ \\
  \hline
  $E_{\rm{kin}}$    & -0.94    & -7.90    && -2.55    & -9.80  &&  0.97     &  -8.18  \\
  $E_\sigma^D+E_\omega^D$ & 9.24 & 13.26  && 15.48    & 17.91  &&  7.43     &  12.20 \\
  $E_\sigma^E+E_\omega^E$ & -0.34 & -0.24 && -1.65    & -0.49  &&  0.00     &   0.00  \\
  $E_{\rm{others}}$ & -1.76    &  0.73    && -1.42    &  0.92  && -0.92     &   0.77 \\
  \hline
  \multirow{2}{*}{Sum}    &  6.20    &  5.85    &&  9.86    &  8.54  &&  7.48     &   4.79 \\
  \cline{2-3} \cline{5-6} \cline{8-9}
                    & \multicolumn{2}{c}{12.05} && \multicolumn{2}{c}{18.40} && \multicolumn{2}{c}{12.27} \\
  \hline\hline
  \end{tabular}
\end{table}

To control the overestimated $E_{\sigma,N\Lambda}+E_{\omega,N\Lambda}$ in RHF models, therefore, a reduction to the meson-hyperon coupling strength $g_{\sigma\Lambda}$ is necessary, correspondingly an alternation to the Dirac effective mass $M_\Lambda^*=M_\Lambda+\Sigma_{S,\Lambda}$. From PKO1-$\Lambda1'$ to PKO1-$\Lambda1$, the value of $g_{\sigma\Lambda}/g_{\sigma N}$ is slightly weakened, also as the case for other RHF functionals in Table.\ref{Tab:Coupling Constants}, to rebalance the effective nuclear force with the strangeness degree of freedom. Within PKO1-$\Lambda1$, the contribution from $E_{\sigma,N\Lambda}+E_{\omega,N\Lambda}$ is then dropped down to $-12.86$ MeV so as to get a reasonable description of the total binding energy for single-$\Lambda$ hypernuclei, consequently the $\Lambda$ separation energy in accord with the experimental data as well, as seen in Table.\ref{Tab:Binding Energies}.

\subsection{Local self-energies and $\Lambda$ spin-orbit splitting}

\begin{figure}[htbp]
  \centering
  \includegraphics[width=0.45\textwidth]{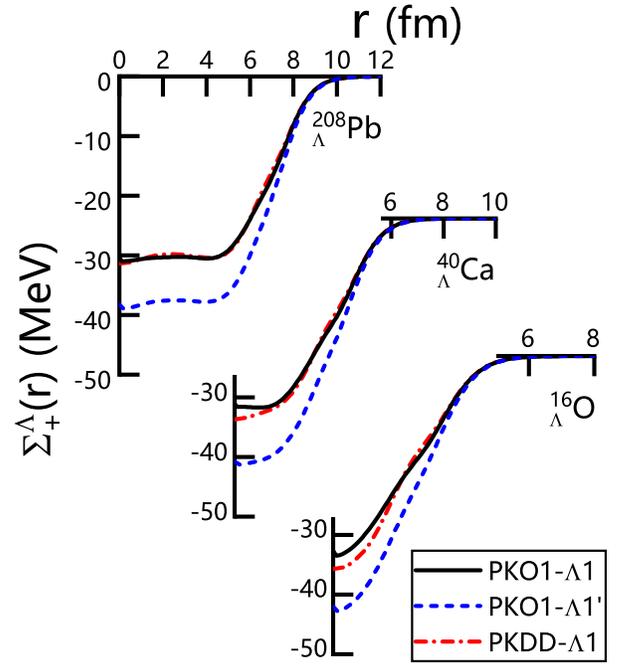}
  \caption{(Color Online) The $\Lambda$ local self-energy $\Sigma_+^\Lambda(r)$ for the single-$\Lambda$ hypernuclei $^{16}_\Lambda$O, $^{40}_\Lambda$Ca and $^{208}_\Lambda$Pb. The results are calculated by the RHF effective interactions PKO1-$\Lambda1$, PKO1-$\Lambda1'$ and the RMF one PKDD-$\Lambda1$.}\label{Fig:VaddS}
\end{figure}

As the $\Lambda$ binding energies have been reproduced well in the selected RHF models, it is now worthwhile to have a look at the $\Lambda$'s single-particle properties such as its self-energies and the energy levels, to illustrate further the influence of the Fock diagram on $\Lambda$-involved nuclear physics. In Fig.\ref{Fig:VaddS}, the local $\Lambda$ self-energies $\Sigma_+^\Lambda$ are exhibited for the single-$\Lambda$ hypernuclei $^{16}_\Lambda$O, $^{40}_\Lambda$Ca and $^{208}_\Lambda$Pb. It is found that PKO1-$\Lambda1'$ gives deeper single-particle potentials than PKDD-$\Lambda1$ when keeping the same ratio of $g_{\sigma\Lambda}/g_{\sigma N}$. With the decreased coupling of $g_{\sigma\Lambda}$, the distributions of $\Sigma_+^\Lambda$ in PKO1-$\Lambda1$ become comparable with ones in PKDD-$\Lambda1$ again, in consistence with the trends of the $\Lambda$ binding energies discussed above. The evolved feature of the equilibrium of nuclear dynamics is represented then by the dependence of the $\Lambda$ self-energy on both in-medium couplings of the meson-nucleon $g_{\phi N}$ and the meson-hyperon $g_{\phi \Lambda}$, as seen in Eqs.\eqref{eq:Sigma_S,B} and \eqref{eq:Sigma_0,B}.

\begin{figure}[hbpt]
  \centering
  \includegraphics[width=0.48\textwidth]{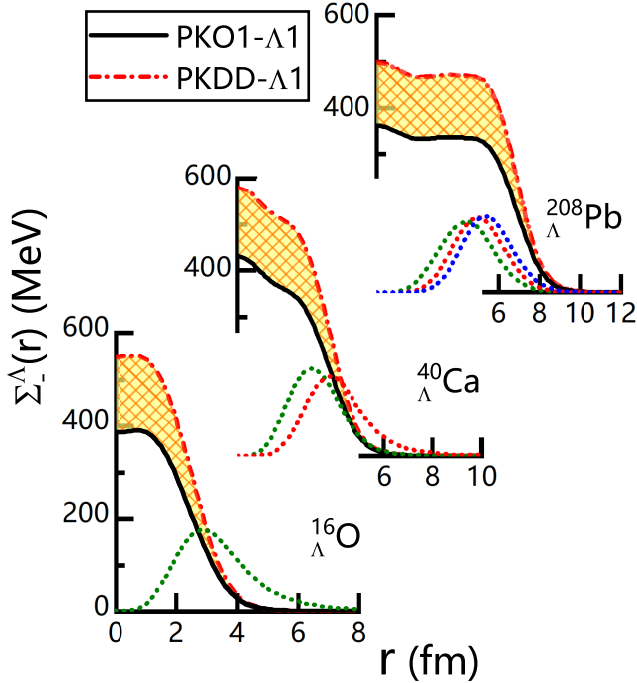}
  \caption{(Color Online) The $\Lambda$ local self-energy $\Sigma_-^\Lambda(r)$ for the single-$\Lambda$ hypernuclei $^{16}_\Lambda$O, $^{40}_\Lambda$Ca and $^{208}_\Lambda$Pb. The results are calculated by the RHF effective interaction PKO1-$\Lambda1$ in comparison with the RMF one PKDD-$\Lambda1$, where the grid pattern denotes their difference. The dotted lines represent the radial density distributions of $\Lambda$ states $1p_{3/2}$ (green), $1d_{5/2}$ (red) and $1f_{7/2}$ (blue) within PKO1-$\Lambda1$.}\label{Fig:VminusS}
\end{figure}

Although the similar $\Sigma_+^\Lambda$ is given by the selected RHF and RMF models, the values of the local self-energy $\Sigma_-^\Lambda$ within PKO1-$\Lambda1$ are obviously smaller than those in PKDD-$\Lambda1$, as well as its radial slope around the surface of hypernuclei, which are plotted in Figure.\ref{Fig:VminusS} for $^{16}_\Lambda$O, $^{40}_\Lambda$Ca and $^{208}_\Lambda$Pb, respectively. When the Fock diagram of $NN$ and $\Lambda\Lambda$ is introduced in CDF approaches, the $N\Lambda$ coupling changes simultaneously. Therefore, the agreements of $\Sigma_+^\Lambda$ within RHF and RMF have no choice but to cause a considerable deviation of $\Sigma_-^\Lambda$, highlighted by the yellow grid pattern in the figure, due to the opposite sign of the isoscalar $\Sigma_{S,\Lambda}$ appeared in $\Sigma_+^\Lambda$ and $\Sigma_-^\Lambda$. With the suppressed $g_{\sigma\Lambda}$ couplings, the RHF functionals then give smaller $\Sigma_-^\Lambda$ than RMF one.
From Eq.\eqref{eq:Vso}, the spin-orbit coupling potential of $\Lambda$ hyperon is determined actually by the radial derivative of $M_+$, correspondingly of $\Sigma_-^\Lambda$, so it is expected that the discrepancy of $\Sigma_-^\Lambda$ with different CDF functionals affects explicitly their performance in describing the $\Lambda$ spin-orbit splitting.

\begin{figure}[hbpt]
  \centering
  \includegraphics[width=0.48\textwidth]{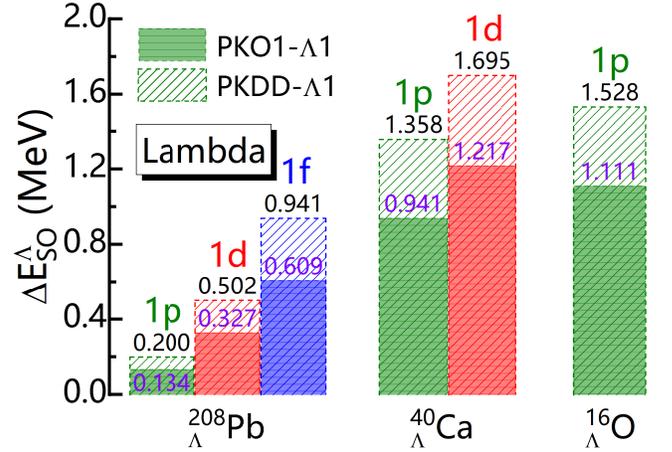}
  \caption{(Color Online) The spin-orbit splittings of $\Lambda$ spin partner states $1p$, $1d$, and $1f$ for the ground state of single-$\Lambda$ hypernuclei $^{16}_\Lambda$O, $^{40}_\Lambda$Ca and $^{208}_\Lambda$Pb. The results are calculated by the RHF effective interaction PKO1-$\Lambda1$ (values in black) in comparison with the RMF one PKDD-$\Lambda1$ (values in violet).}\label{Fig:ESO}
\end{figure}

The hyperon's spin-orbit splitting can be estimated by the difference of $\Lambda$ single-particle energies between a couple of spin partner states, which is
\begin{align}
\Delta E_{\rm{SO}}^\Lambda \equiv \varepsilon_{j_\Lambda=l_\Lambda-1/2} - \varepsilon_{j_\Lambda=l_\Lambda+1/2}.
\end{align}
According to the Schr\"odinger-like equation in Eq.\eqref{eq:SchLike}, $\Delta E_{\rm{SO}}^\Lambda$ is mainly correlated with the spin-orbit coupling potential $V_{\rm{SO},\Lambda}$ since the spin partners contribute the similar values to other terms\cite{liang_hidden_2015}. As much smaller in magnitude than the local terms, the influence of the nonlocal self-energies of hyperon on $V_{\rm{SO},\Lambda}$ are ignored in the following discussions. Taking the ground state of $^{16}_\Lambda$O, $^{40}_\Lambda$Ca and $^{208}_\Lambda$Pb as examples, the calculated spin-orbit splittings for the spin partners $1p$, $1d$, and $1f$ of $\Lambda$ hyperon are depicted in Fig.\ref{Fig:ESO}.

It is clear in Fig.\ref{Fig:ESO} that the $\Lambda$ spin-orbit splitting given by the RHF functional PKO1-$\Lambda1$ is systematically lower than the RMF's PKDD-$\Lambda1$, complying with the fact that in the RHF case there are smaller values of $\Sigma_-^\Lambda$ and effective spin-orbit coupling potential $V_{\rm{SO},\Lambda}$. Thus, the inclusion of relativistic Fock diagram in the CDF framework, which alters the equilibrium of nuclear dynamics, impacts enormously the single-particle properties of hypernuclei. In addition, it is seen that the values of $\Delta E_{\rm{SO}}^\Lambda$ for $\Lambda$'s $1p$ partners decrease gradually from $^{16}_\Lambda$O to $^{208}_\Lambda$Pb in both RHF and RMF cases, which could be explained by plotting the radial density distributions of $\Lambda$'s $1p$ states as shown in green dotted lines in Fig.\ref{Fig:VminusS}. In fact, the upper components of the radial wave functions, which dominate the density profile, are nearly identical to each other in the spin partner states\cite{zong_relativistic_2018}, so the values are just given for $1p_{3/2}$. From $^{16}_\Lambda$O to $^{208}_\Lambda$Pb, the density peak is found to move from the surface to the interior of the hypernucleus, where the radial slope of $\Sigma_-^\Lambda$ evolves gradually so that $V_{\rm{SO},\Lambda}$ of $\Lambda$ hyperon drops down correspondingly. Besides, a similar analysis is suitable for the evolution of $\Delta E_{\rm{SO}}^\Lambda$ in the spin partner states of $^{208}_\Lambda$Pb, where the increase of $\Delta E_{\rm{SO}}^\Lambda$ from $1p$ to $1f$ is related to the density peak approaching closely to the surface.

It is generally believed that the energy splitting between spin partner states in single-$\Lambda$ hypernuclei is very small in comparison with that for nucleons\cite{may_observation_1981, may_observation_1983, ajimura_observation_2001, Akikawa_Hypernuclear_2002, Kohri_hypernucleu_2002}. From the obtained RHF functionals here, it is still difficult to give a comparable value to the experimental or empirical data although quenching effects in $\Lambda$ spin-orbit splitting are already gained in RHF calculations, e.g., $\Delta E_{\rm{SO}}^\Lambda\thickapprox 1.181$ MeV in PKO1-$\Lambda1$ for $1p$ partners of $^{13}_\Lambda$C. Theoretically, there exist several mechanisms to reduce the predicted $\Delta E_{\rm{SO}}^\Lambda$ such as the SU(3) symmetry breaking and the tensor coupling\cite{NOBLE_potential_1980, jennings_dirac_1990}, in CDF approaches which could be involved by changing their effective interactions. In fact, it is found that a good linear correlation exists in RMF models between two ratios $g_{\sigma\Lambda}/g_{\sigma N}$ and $g_{\omega\Lambda}/g_{\omega N}$\cite{Keil_hypernuclei_2000, Wang_Lambda_2013, Rong_RMF_2021}. Thus, the single-particle properties of $\Lambda$-hypernuclei could be adjusted by evolving the meson-hyperon coupling strengths while maintaining the well-reproduced bulk properties.

\begin{table}[hbpt]
  \caption{The series of CDF effective interactions obtained by alternating the $\omega$-$\Lambda$ coupling constant $g_{\omega\Lambda}/g_{\omega N}$ from 0.3 to 0.8 (for convenience marked by the suffix ``x"). The $\sigma$-$\Lambda$ couplings $g_{\sigma\Lambda}/g_{\sigma N}$ are fitted by minimizing the root-mean-square deviation $\Delta$ (in MeV) from the experiment values of $\Lambda$ binding energies of $^{16}_\Lambda$O, $^{40}_\Lambda$Ca and $^{208}_\Lambda$Pb.}
  \label{Tab:different ome}
  \renewcommand{\arraystretch}{1.5}
  \setlength{\tabcolsep}{2pt}
  \begin{tabular}{cccccccc}
  \hline\hline
    & $g_{\omega\Lambda}/g_{\omega N}$ & 0.3  &  0.4  &  0.5  &  0.6  &  0.7  &  0.8 \\
    \hline
    \multirow{2}{*}{PKO1-$\Lambda$x} & $g_{\sigma\Lambda}/g_{\sigma N}$ &0.334  &0.405  &0.477  &0.549  &0.621  &0.692      \\
    & $\Delta$                &0.936  &0.706  &0.495  &0.324  &0.263  &0.356      \\\hline
    \multirow{2}{*}{PKO2-$\Lambda$x} & $g_{\sigma\Lambda}/g_{\sigma N}$ &0.334  &0.404  &0.474  &0.545  &0.614  &0.685      \\
    & $\Delta$                &0.970  &0.746  &0.534  &0.368  &0.240  &0.289      \\\hline
    \multirow{2}{*}{PKO3-$\Lambda$x} & $g_{\sigma\Lambda}/g_{\sigma N}$ &0.331  &0.403  &0.475  &0.546  &0.618  &0.690      \\
    & $\Delta$                &0.735  &0.543  &0.400  &0.362  &0.447  &0.599      \\\hline
    \multirow{2}{*}{PKDD-$\Lambda$x} & $g_{\sigma\Lambda}/g_{\sigma N}$ &0.322  &0.403  &0.485  &0.566  &0.647  &0.729      \\
    & $\Delta$                &1.150  &0.932  &0.709  &0.487  &0.280  &0.162      \\
  \hline\hline
  \end{tabular}
\end{table}

\begin{figure}[hbpt]
  \includegraphics[width=0.48\textwidth]{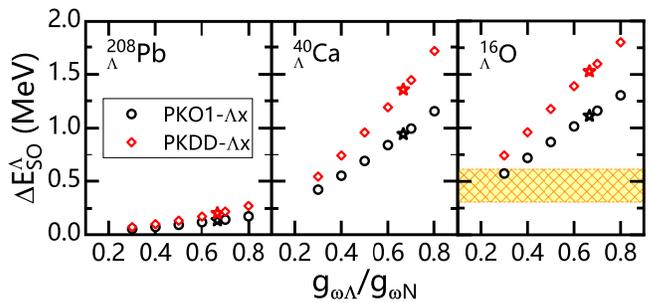}
  \caption{(Color Online) The spin-orbit splittings of $\Lambda$ spin partner state $1p$ for the single-$\Lambda$ hypernuclei $^{16}_\Lambda$O, $^{40}_\Lambda$Ca and $^{208}_\Lambda$Pb, evolving with $\omega$-$\Lambda$ coupling strength $g_{\omega\Lambda}/g_{\omega N}$. The results are calculated by the RHF effective interaction series PKO1-$\Lambda$x in comparison with the RMF one PKDD-$\Lambda$x taken from Tab.\ref{Tab:different ome}, while the empirical value for $^{16}_\Lambda$O shown with the grid pattern is taken from Ref. \cite{motoba_parity-mixing_1998}. The starred points are the cases at $g_{\omega\Lambda}/g_{\omega N} = 0.666$.}\label{Fig:ESO with Omega}
\end{figure}

To follow this idea, we carried out the fitting procedure of $\Lambda$-relevant parameters again but release the constraint on the vector coupling $g_{\omega\Lambda}/g_{\omega N}$, varying from 0.8 to 0.3. Then the scalar coupling strength $g_{\sigma\Lambda}/g_{\sigma N}$ is determined by reproducing the experimental $\Lambda$ binding energies of hypernuclei in the same way as $\Lambda1$ in Table.\ref{Tab:Coupling Constants}. The obtained $\Lambda$x series of CDF functionals for the single-$\Lambda$ hypernuclei are listed in Table \ref{Tab:different ome}. With decreasing ratio of $g_{\phi\Lambda}/g_{\phi N}$, the root-mean-square deviation $\Delta$ raises to a certain extent, resulting in somewhat worse predictions to the $\Lambda$ separation energies. Focusing on their role in $\Lambda$ single-particle properties, these series of CDF functionals can be used to evolve the hyperon's spin-orbit splitting. The results for $\Lambda$ spin partners $1p$ of $^{16}_\Lambda$O, $^{40}_\Lambda$Ca and $^{208}_\Lambda$Pb are shown in Fig.\ref{Fig:ESO with Omega}.

From Fig.\ref{Fig:ESO with Omega}, it is clear that the $\Lambda1p$ spin-orbit splitting goes down enormously as the meson-hyperon coupling strength $g_{\omega\Lambda}/g_{\omega N}$ decreases, especially for the light hypernucleus $^{16}_\Lambda$O. In comparison with PKDD-$\Lambda$x, PKO1-$\Lambda$x predicts smaller $\Delta E_{\rm{SO}}^\Lambda$ at the same value of $g_{\omega\Lambda}/g_{\omega N}$. From the literature, the value of $\Lambda1p$ splitting for $^{16}_\Lambda$O is estimated empirically around $300\le\Delta E_{\rm{SO}}^\Lambda\le 600$ keV\cite{motoba_parity-mixing_1998}. As seen in the figure, PKO1-$\Lambda$x enter into such an area (the yellow grid pattern) earlier than PKDD-$\Lambda$x when $g_{\omega\Lambda}/g_{\omega N}$ decreases. Thus, to comply with the empirical constraint the RHF models give a larger parameter space of meson-hyperon couplings, which could be $g_{\omega\Lambda}/g_{\omega N}\lesssim0.3$ and $g_{\sigma\Lambda}/g_{\sigma N}\lesssim0.334$ for PKO1 functionals. Moreover, it is manifested that the predicted splitting from two functionals would approach each other when the meson-hyperon coupling strengths $g_{\phi\Lambda}$ weaken, which is attributed to the fact that the hyperon-induced mean field and correspondingly the self-energy $\Sigma_-^\Lambda$ is eliminated with decreasing coupling, and so is its disparity between different functionals.

\section{Summary}\label{Summary and Outlook}
In summary, the relativistic Hartree-Fock (RHF) theory has been extended to include the degree of freedom of $\Lambda$ hyperon. Several $\Lambda$-nucleon effective interactions are introduced with density-dependent meson-hyperon couplings, as their strengths are determined by fitting $\Lambda$ separation energies to the experimental data for single-$\Lambda$ hypernuclei. Focusing on the $\Lambda$-involved effects, the obtained RHF functionals are adopted to study the role of Fock diagram from a viewpoint of the equilibrium of nuclear dynamics in $\Lambda$ hypernuclei, in comparison with the RMF calculation.

In the case of single-$\Lambda$ hypernuclei, $\Lambda$ contributes to the isoscalar potential energy from both Hartree and Fock channels, dominated by $E_{\sigma,\Lambda N}+E_{\omega,\Lambda N}$ rather than $E_{\sigma,\Lambda\Lambda}+E_{\omega,\Lambda\Lambda}$ since only one hyperon exists. Demonstrated by comparing a relative ratio $\chi_\Lambda$ of hyperon to $\chi_N$ in nucleon channel, the equilibrium of nuclear dynamics described by the RHF models in single-$\Lambda$ hypernuclei is clearly deviated from that in normal atomic nuclei, indicating a different role of Fock terms via $\Lambda$ hyperon from the nucleons. As a consequence of $\Lambda$ implantation, being the overwhelmed $\Lambda N$ attraction via the isoscalar Hartree over the $\Lambda\Lambda$ exchange terms, RHF models then ask a systematically reduced $\sigma$-$\Lambda$ coupling strength $g_{\sigma\Lambda}$ as compared to RMF's case so as to rebalance the strangeness-bearing effective nuclear force.

Turning to the single-particle properties of $\Lambda$ hypernuclei, it is found that the selected RHF functionals give relatively smaller values of the $\Lambda$ local self-energy $\Sigma_-^\Lambda$ and the spin-orbit coupling potential $V_{\rm{SO},\Lambda}$ than RMF one PKDD-$\Lambda1$, due to the reduced $\sigma$-$\Lambda$ coupling. As a result, these RHF models predict correspondingly a quenched $\Lambda$ spin-orbit splitting in comparison with the RMF case, examples shown for $^{16}_\Lambda$O, $^{40}_\Lambda$Ca and $^{208}_\Lambda$Pb. Our work confirms that the inclusion of Fock diagram in a covariant energy density functional plays an essential and non-negligible role indeed in understanding the origin of why the $\Lambda$ spin-orbit splitting in hypernuclei is very small in comparison with that of nucleons.

Finally, a possible way to reduce the uncertainty of such an issue is discussed, to reproduce the splitting $\Delta E_{\rm{SO}}^\Lambda$ theoretically within the selected RHF models. By evolving the hyperon-relevant couplings $g_{\sigma\Lambda}$ and $g_{\omega\Lambda}$ simultaneously, inspired by the discovered linear correlation between them, the predicted $\Lambda$ spin-orbit splitting could decrease efficiently. As compared to PKDD-$\Lambda1$, the RHF models declare a larger parameter space of meson-hyperon couplings in order to give a complied description with the empirical energy splitting $\Delta E_{\rm{SO}}^\Lambda$ for $\Lambda1p$ partners of $^{16}_\Lambda$O.

It has been pointed out that the tensor coupling embedded in RHF approach plays an important role in treating the delicate balance in the dynamic nuclear medium and controlling the single-particle characters\cite{long_shell_2007, geng_pseudospin_2019}. It is possible to introduce the $\rho$- or $\omega$-tensor couplings in both $NN$ and $N\Lambda$ channels in a self-consistent way within RHF, and to check the induced effect on the $\Lambda$ spin-orbit splitting, which is still ignored in this work and would be considered later on. Moreover, the exchange of strangeness-bearing mesons becomes allowed when the Fock diagrams are involved beyond the relativistic Hartree ones. Since we are studying the single-$\Lambda$ hypernuclei, the contribution of these strangeness-bearing mesons to the effective $\Lambda N$ interaction is expected to be relatively smaller than those of $\sigma$ and $\omega$. However, when investigating multi-$\Lambda$ hypernuclei or infinite hypernuclear systems in which multi hyperons coexist, such strangeness-bearing mesons couplings may not be negligible and need consider seriously. Hence, the model is expected to be developed further by including strangeness-bearing meson exchange for the cases of multi-$\Lambda$ hypernuclei and hyperon star physics.

\begin{acknowledgements}
This work was partly supported by the Natural Science Foundation of China under Grant No. 11875152, the Strategic Priority Research Program of Chinese Academy of Sciences under Grant No. XDB34000000, and the Fundamental Research Funds for the Central Universities under Grant No. lzujbky-2021-sp36. Shi Yuan Ding and Zhuang Qian contributed equally to this paper. In addition, they thank Jing Geng for helpful discussions.
\end{acknowledgements}


\end{document}